\documentclass[11pt]{article}
\usepackage{graphicx,pstricks}
\usepackage{lscape}
\usepackage{amsmath}
\title{\LARGE \bf Quantum Violation of Bell's Inequality: \\ a misunderstanding based on \\ a mathematical error of neglect}

\author{{\large Frank Lad,\vspace{.5cm} email: frank.lad@canterbury.ac.nz}\\
{\normalsize{\it Department of Mathematics and Statistics, University of Canterbury}}}

\date{29 March 2020}

\begin{document}

\maketitle

\begin{abstract}

The fabled violation of Bell's inequality by the probabilistic specifications of quantum
mechanics is shown to derive from a mathematical error. The inequality, designed to assess consequences of Einstein's principle of local realism, pertains to four polarization products on the same pair of photons arising in a gedankenexperiment. The summands of the CHSH  quantity  $s(\lambda)$ inhere four symmetric functional relations which have long been neglected in analytic considerations.  Its expectation $E[s(\lambda)]$  is not the sum of four ``marginal'' expectations from a joint distribution, as quantum theory explicitly avoids such a specification. Rather, $E[s(\lambda)]$ has four distinct representations as the sum of three expectations of polarization products plus the expectation of a fourth which is restricted to equal a function value determined by the other three.  Analysis using Bruno de Finetti's fundamental theorem of prevision (FTP) yields only a bound for $E(s)$ within $(1.1213, 2]$, surely not $2\sqrt{2}$ at all.  The 4-D polytope of cohering joint $P_{++}$ probabilities at the four stipulated angle settings are displayed as passing through 3-D space.  Aspect's ``estimation'' is based on polarization products from different photon pairs that do not have embedded within them the inhering functional relations.   When you do actively embed the restrictions into Aspect's estimation procedure, it yields an estimate of 1.7667, although this is not and cannot be definitive.

\end{abstract}


\section{Introduction}

As surprising as this may sound, claims that probabilistic specifications of
quantum mechanics defy the mathematical prescription known as Bell's inequality are just
plain wrong.  This may be difficult for you to accept, depending on how wedded you are to
the outlook that gives rise to them.  You will not be alone.  The eminent journal {\it Nature}
(2015, {\bf 526}, 649-650) flamboyantly announced to its readership the
``Death by experiment for local realism'' as an introduction to its
publication of experimental results achieved at the Technical University of Delft.  These were
proclaimed to have closed simultaneously all seven loopholes that had been suggested as
possible explanations of the touted violations of the inequality.  In the eyes of the
professional physics community, the matter is now closed.  My claim is that
the touted violation of the inequality derives from a mathematical mistake, an error of neglect.
Its recognition relies only on a basic understanding of functions of many variables and on
standard features of applied linear algebra. This presentation is designed for any sophisticated
reader not put off by equations per se, who has followed this issue at least at the level of popular description of scientific activity.\\

It is clear in his own writings that John Bell himself
was puzzled by the implications of his inequality (1964, 1966, 1971, 1987).  He suspected that something was
wrong with the understanding that the probabilities of quantum mechanics seem to defy its
 structure, and he expressed undying confidence
 that this error would be discovered in due time.  I am making a bold claim that I
have found the error he sought.  I accept all probabilistic assertions supported by quantum theory,
and I exhibit their implied support of the inequality bounds.  \\

I do not contest the experimental results of the Delft group, nor any of the related
experimentation which has followed from the pathbreaking initial work of Alain Aspect.  I do
contest the inferences they are purported to support.
In this note I will first review the derivation of the inequality in the context to which it
applies, featuring its relation to Einstein's principle of local realism.  The review will
focus on the CHSH form of the inequality to which Aspect's optical experimentation is
considered to be relevant.  Identifying the neglected functional relations that are
involved in a thought experiment on a single pair of photons,
I will show why the claims to defiance of the inequality are mistaken, and
how to derive the actual implications of quantum theory for the probabilities
under consideration.  Further will be shown
why Aspect's computations (and all subsequent extensions)
proposed to exhibit empirical confirmation of
the inequality defiance are ill considered, and how they ought to be adjusted.
This demonstration relies on the
computational mechanics of Bruno
de Finetti's fundamental theorem of probability.  The results
are displayed both algebraically and geometrically.

\section {The physical setup of four experiments providing context \\
\hspace*{.6cm}for Bell's inequality in CHSH form: a 16-D problem}

We shall review the setup of an optical variant of Bell's experiment, designed by Alain Aspect
in the 1980's to take advantage of a formulation of the problem proposed by Clauser, Horne,
Shimony, and Holt (CHSH, 1969). The original discussions of the inequality violation
were couched in terms of observations of spins of
paired electrons.  Although specific algebraic details differ for the two types of
experimental situation, the conclusions reached would be identical. \\

An experiment is conducted on a pair of photons traveling in opposite directions along an
axis, ${\bf z}$, from a common source.  The direction one of the photons travels toward detector $A$ on
the left is directly opposite to the direction its paired photon travels toward detector $B$ on the
right: $\,{\bf z}_A = -{\bf z}_B\,$.
At the end of their respective journeys, each of the paired photons engages polarising material
that either allows it to pass through or to be deflected.  \\
\begin{figure}[!h]
\begin{center}
\hspace*{.75cm} \includegraphics[width=0.6\linewidth]{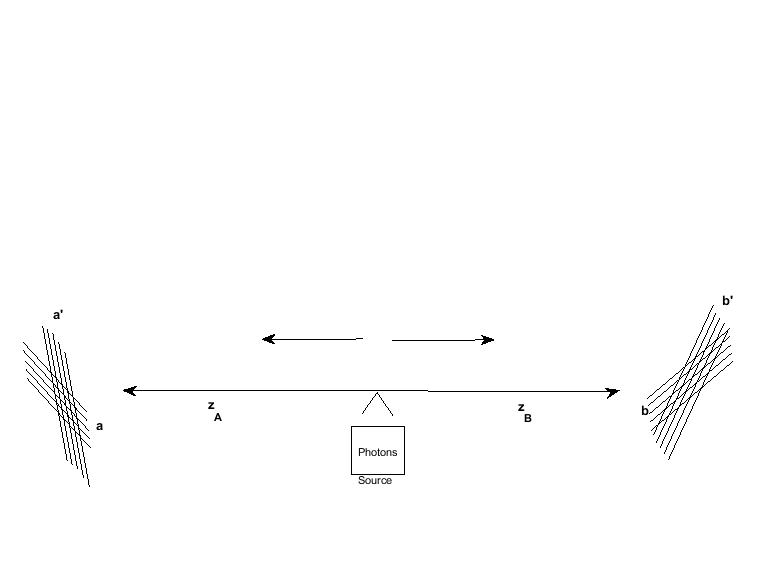}\\
\caption{Polarizing material is aligned at the detection stations of $A$ and $B$, each with two possible
choices of direction in the $(x,y)$ dimension relative to the ${\bf z}$ direction of the incoming photon:
direction ${\bf a}$ or ${\bf a'}$ at station $A$, and ${\bf b}$ or ${\bf b'}$ at station $B$.}
\label{fig:photondirectionsetup}
\end{center}
\end{figure}\\
The detection of a photon that passes
through the polarizer is designated by denoting the numerical value of $A = +1$, while the detection
of a photon as blocked
is designated by the value of $A = -1$.  The polarizer addressed by photon $A$ is directed at a
variable angle
${\bf a^*}$  in the $(x,y)$ plane perpendicular to $\,{\bf z}_A$.  This polarization direction
can be set in either of two specific
angles designated as ${\bf a}$ and ${\bf a'}$ in the experimental setups we shall consider.
Similarly, the direction of the polarizer
met by the photon at station $B$ can be set at either angle ${\bf b}$ or ${\bf b'}$ in its $(x,y)$ plane.
Depending on the specific pair of polarization angles ${\bf a^*}$,${\bf b^*}$  chosen for any particular
experiment, we shall observe the paired values of either $(A({\bf a}),B({\bf b})),\ $ or $(A({\bf a}),B({\bf b'})),\ $ or $(A({\bf a'}),B({\bf b}))$, or $(A({\bf a'}),B({\bf b'}))$.  Since the observations of the $A$ and the $B$ photon
detections can each
equal either $+1$ or $-1$ whatever the angle pairing might be, the chosen observation pair $(A({\bf a^*}),B({\bf b^*}))$
will equal one of the four possibilities $(+,+), (+,-), (-,+)$, or $(-,-)$, where we are suppressing here the
needless numeric values of $1$ in each designated pairing.\\

Experimental choices of the two polarization angle directions yield a specific {\it relative angle}
between them at $A$ and $B$ in any given experiment.  Using Aspect's notation that
parentheses around a pair of directions denotes the relative angle between them, the
experimental detection angle settings  $({\bf a}^*,{\bf z}_A)$
and $({\bf b}^*,{\bf z}_B)$ imply the {\it relative} angle between polarizers at stations $A$ and $B$
in the $(x,y)$ dimension as $({\bf a}^*,{\bf b}^*)$.
Bell's inequality is relevant to this context in which the two photon
polarization directions can be paired at any one of four distinct relative angles,
denoted by the parenthetic pairs $({\bf a},{\bf b})$,  $({\bf a},{\bf b'})$,  $({\bf
a'},{\bf b})$, or \ $({\bf a'},{\bf b'})$.\\

In order to view the relative angles we are talking about, mentally we would need to swing
the $(x,y)$ plane as it is viewed by the photon directed to station $A$ around by $180^\circ$
and superimpose it on the $(x,y)$ plane as it is viewed by the photon directed to station $B$.
In this manner we can understand the size and meaning of the relative angles {\it between} the
various values of polarization orientations ${\bf a^*}$  and ${\bf b^*}$ as seen here in Figure $2$.
\begin{figure}[!h]
\begin{center}
\includegraphics[scale=0.6]{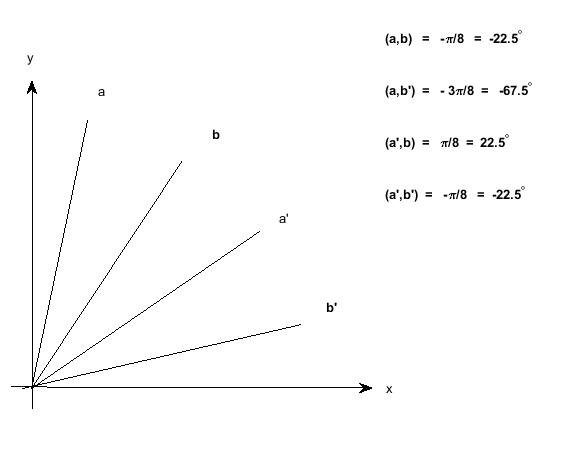}
\caption{Directional vectors of the polarization angle settings at the observation
stations $A$ and $B$, viewed in a common axis orientation.  The specific relative angle
size settings displayed are the extreme violation values, a feature to be discussed.}
\label{fig:xtrmviolationanglesFig2}
\end{center}
\end{figure}\\

The theory of quantum mechanics motivates specification of probabilities for the four observable
outcome possibilities of the polarization experiment as depending on the {\it relative angle} $({\bf a}^*,{\bf b}^*)$
between the direction vectors of the polarizers at stations $A$ and $B$.  For any such relative angle
pairing, the probabilities specified by quantum theory for the four possible experimental observations
$\{++, +-, -+, --\}$ are
\begin{center}
$P[(A({\bf a}^*) = +1)(B({\bf b}^*) = +1)] \ = \ P[(A({\bf a}^*) = -1)(B({\bf b}^*) =
-1)] \ = \ \frac{1}{2}\;cos^2({\bf a}^*,{\bf b}^*)\;$,\vspace{.2cm}\ \ \ and\\
\hspace*{.33cm}$P[(A({\bf a}^*) = +1)(B({\bf b}^*) = -1)] \ = \ P[(A({\bf a}^*) =
-1)(B({\bf b}^*) = +1)] \ = \ \frac{1}{2}\;sin^2({\bf a}^*,{\bf b}^*)\;$. \ \ \ \ \ $(1)$
\end{center}
For efficiency in what follows, we shall denote the four probabilities appearing
in equations $(1)$ by $P_{++}, P_{--}, P_{+-}$, and $P_{-+}\,$ when the pertinent angle
setting is evident. \\

These four probabilities surely sum to equal $1$, because the sum of $cos^2 + sin^2$ of
any angle equals $1$.   A few properties of the joint probability mass function (pmf) they
compose should be noticed.  Firstly, the four probabilities can be specified
by the value of any one of them.  The equations $(1)$ stipulate that
no matter what the relative angle $({\bf a}^*,{\bf b}^*)$
may be, the values of $P_{++}=P_{--}$, and $P_{-+}=P_{+-}$.  Since the four probabilities
do sum to $1$ then, the specification of $P_{++}$ as the value $p$, for example, implies
that the pmf vector
$[P_{++}, P_{--}, P_{+-}, P_{-+}]\,$ would be $[p, p, (1-2p)/2, (1-2p)/2]$.
Another implication of this feature is that the probabilities for the paired detection outcomes depend
only on the {\it product} of the two measurements.  For
both outcomes $++$ and $--$ yield a product of $+1$ and both outcomes $+-$ and $-+$ yield a product of $-1$.
Thus, the QM-motivated distribution for the experimental value of the polarization {\it product}
$A({\bf a^*})B({\bf b^*})$ is specified by
$P[A({\bf a^*})B({\bf b^*}) = +1] \ = \ cos^2({\bf a^*},{\bf b^*})$ \ and \ $P[A({\bf a^*})B({\bf b^*}) = -1] \ = \ sin^2({\bf a^*},{\bf b^*})$. \\

As will be important to recognize in what follows,  the expected value (first moment) of this distribution
for the detection product is \vspace{.4cm}\\
\hspace*{3cm}$E[A({\bf a^*})B({\bf b^*})] \ = \ (+1)cos^2({\bf a^*},{\bf b^*}) + (-1)sin^2({\bf a^*},{\bf b^*})\vspace{.3cm} \\
\hspace*{5.65cm} =  \ cos^2({\bf a^*},{\bf b^*}) - sin^2({\bf a^*},{\bf b^*}) \ = \ cos\,2({\bf a}^*,{\bf
b}^*)\;$ \vspace{.4cm} \ \ \ \ \ \ \ \ \ \ \ \ \ \ (2) \\
according to
standard double angle formulas.  It is worthwhile reminding right here that ``the expected value of a probability
distribution'' is the ``first moment''
of the distribution.  Geometrically, it is the point of balance of the probability mass function weights when they are
positioned in space at the places where the possible observations to which they pertain might occur.  It is a property
of a probability distribution for
the outcome of a specific single observable variable.  A final peculiarity of equation $(2)$ which will be useful far down the road in this explication is that the expectation value $E[A({\bf a^*})B({\bf b^*})]$ can also be represented\vspace{.4cm} as\\
\hspace*{3cm} $E[A({\bf a^*})B({\bf b^*})] \ \ = \ \ 2\,cos^2({\bf a}^*,{\bf b}^*) - 1 \ \ = \ \ 4\,P_{++}({\bf a^*},{\bf b^*}) -1$\vspace{.4cm}\ .  \ \ \ \ \ \ \ \ \ \ \ \ \ $(3)$\\
\noindent For the value of $sin^2({\bf a^*},{\bf b^*})$ appearing in the final line of equation  $(2)$
can also be written as $1-cos^2({\bf a^*},{\bf b^*})$.  Enough of this for now. \\

 Secondly, again
no matter what the relative angle $({\bf a}^*,{\bf b}^*)$ may be, the marginal probabilities that
the detection observation of the photon equals $+1$ at either angle ${\bf a}^*$ or ${\bf b}^*$
is equal to $1/2$.  For the standard margining equation for the result of a paired experiment yields \vspace{.2cm}  \\
\indent \hspace*{.1cm}$P[(A({\bf a}^*) = +1) \ \ = \ \ P[(A({\bf a}^*) = +1)(B({\bf b}^*) = +1)] \ + \ P[(A({\bf a}^*) = +1)(B({\bf b}^*) = -1)] \vspace{.2cm} \\
\hspace*{3.6cm} = \ \  \frac{1}{2}\;cos^2({\bf a}^*,{\bf b}^*)\; + \frac{1}{2}\;sin^2({\bf a}^*,{\bf b}^*)\; \ = \ \ 1/2$. \ \ \ \ \ \ \ \ \ \ \ \ \ \ \ \ \ \ \ \ \ \ \ \ \ \ \ \ $\; $\ \ \ (4)\\

This result codifies a touted feature of physical processes at quantum
scales of magnitude, that the photon behaviours of particle pairs are understood to be {\it entangled}.  Since the probability
for the joint photon behaviour $P[(A({\bf a}^*) = +1)(B({\bf b}^*) = +1)]$ does not factor into the product of their
marginal probabilities $P[(A({\bf a}^*) = +1)]$ and $P[(B({\bf b}^*) = +1)]$, the conditional distribution for
either one of these events depends on the context of the conditioning behaviour:\vspace{.2cm}\\
\hspace*{1.0cm}
$P[(A({\bf a}^*) = +1)|(B({\bf b}^*) = +1)] \ \ = \ \ cos^2({\bf a}^*,{\bf b}^*) \ \ \neq \ \ P[(A({\bf a}^*) = +1) \ \ = \ \ 1/2\vspace{.2cm}$ , \ \ \ \ \ \ \ (5)\\
\hspace*{1.5cm}
and  \ \ $P[(A({\bf a}^*) = +1)|(B({\bf b}^*) = -1)] \ = \ sin^2({\bf a}^*,{\bf b}^*)$\ , \ \ \ \ \ \ which is different still.\\

We have concluded what we need to say at the moment about the prescriptions of quantum theory relevant to
physical quantum behaviour of a single pair of prepared photons. Before proceeding to the specification of Bell's inequality,
we need to address what quantum theory professes {\it not} to say.

\section{The uncertainty principle: what quantum theory disavows }

Made famous as what is called ``Heisenberg's uncertainty principle'', the theory of quantum mechanics explicitly
disavows claims to what might happen in physical situations that are impossible to instantiate.  Here is an example
relevant to the classical scale of everyday observation.  A dairy farmer may choose to
treat a milking cow with injections
of bovine somatatropine (BST) or may choose not to use such treatment.  However, one cannot follow both programs
on the same dairy cow.  One could treat some cows with BST and some other cows without BST, but one cannot both treat
and not treat the same cow with BST.  Now what will be the daily weight of the milk yield from this cow?  No one knows for sure.  Of course you could well assess {\it conditional} distributions for the weight of the cow's milk yield,  conditional on each of the two treatment strategies.  However, a distribution for the joint yields from following both of these strategies on the same cow (which would be impossible) is meaningless.
This is just common sense.
There are well known statistical procedures for studying the yields from alternative treatment strategies, applying one strategy to one group of cows and the other strategy to another group.  This is a classic statistical problem of agricultural statistics codified as problems in the {\it design of experiments}.  This subject does not concern us here just now. \\


The problem of quantum physics relevant to Bell's inequality concerns this very same issue.  We have identified a physical experiment on a pair of photons, polarising the two of them at an array of possible exclusive angle pairings,
$({\bf a},{\bf b})$,  $({\bf a},{\bf b'})$,  $({\bf a'},{\bf b})$, and \ $({\bf a'},{\bf b'})$.  We could perform our
polarization experiment on a specific pair of photons at any one of these angle pairings.  But we cannot perform all
four experiments {\it on the same pair of photons}.  The theory of quantum mechanics recognizes this fact explicitly and loudly!  Not only do the experimental physicists recognize that this cannot be done, but the theoretical algebraic
mechanism that is used to identify the quantum probabilities we have specified in equations $(1)$ embed this
impossibility into its protocol.  I will state how this recognition is embedded into quantum theory, without deriving its application here in complete detail.\\

Relevant to the polarization product detection in the case
we are formalizing,
the theory of quantum mechanics characterizes the situation of a quantum experiment in terms of a two-dimensional
vector which resides in one of several possible states of possibility. Let's call the state vector ${\bf s} = (s_1,s_2)^T$.
We cannot observe what state the photon pair is in without performing a measurement.
The measurement process is characterised
algebraically by a matrix, call it $H$, which operates on the state vector by multiplication. This matrix, on account of its form, identifies
the observation values that might arise when the measurement is performed.  In our case it identifies that the polarization product value we observe, in whatever paired polarization experiment we perform, might equal $+1$ or $-1$.  When this matrix multiplies the state vector in the algebraic form of $H{\bf s}$, the result of the product is a pair of probabilities specified for these two possible observation values. In our polarization experiment on the photon pair observed at $A({\bf a}^*,{\bf b}^*)$ and $B({\bf a}^*,{\bf b}^*)$, these are the probabilities specified in our equation (1).  (These two probabilities have been split in half there, to account for the fact that both of the paired polarization results  $++$ and $--$ yield a product of $+1$, and both
of the paired polarization results  $+-$ and $-+$ yield a product of $-1$.) The matrix $H$ is said to be a ``Hadamard matrix''. \\

At any rate, this is the mathematical formalism by which the probabilistic specifications of quantum theory are
derived for the various possible results of a quantum experiment.  Each observation possibility is characterised by its own matrix $H$.  Since we have four possible experimental designs under consideration, codified by the paired angle settings $({\bf a},{\bf b})$,  $({\bf a},{\bf b'})$,  $({\bf a'},{\bf b})$, and \ $({\bf a'},{\bf b'})$, there are four distinct matrices, denoted by  $H_{({\bf a},{\bf b})}$,  $H_{({\bf a},{\bf b'})}$,  $H_{({\bf a'},{\bf b})}$, and \ $H_{({\bf a'},{\bf b'})}$ which codify our experimental measurement possibilities.\\

Now, is it possible to perform {\it two} observational measurements on a quantum experimental situation?  The answer is ``in some cases yes, and in some cases no!''  Happily, there is a very simple way to determine whether two distinct measurements can be performed on the same situation of the state of the photons.  Algebraically, a measurement codified by a matrix $H$ is compatible with a simultaneous second measurement $G$ on the same experimental situation if and only if the product of the two matrices commutes!  That is to say, if and only if the products of the matrices are identical no matter which be the order of the multiplication: \ $HG = GH$.  To check whether this is true or not is
a simple matter of performing  the algebra of multiplying the matrices.  If so, the two matrices are said to be ``Hermitian''.\\

An example of commuting operators has already been broached, without my having mentioned it.
There is an operator matrix $H_{{\bf a}}$ that codifies
the detection observation of the photon at station $A$ with polarization direction ${\bf a}$, and another which codifies a detection observation at station $B$ with polarization direction ${\bf b}$.  Call it $H_{{\bf b}}$.  Now it is a mere matter of mathematical derivation to find that the product of these
two operator matrices does not depend of which one multiplies the other:  that is, $H_{{\bf a}}H_{{\bf b}}$ = $H_{{\bf b}}H_{{\bf a}}$.  This lets us know formally that indeed we can measure the detection of the two photons at both of the stations $A$ and $B$.  That is why we denote this operator by $H_{({\bf a},{\bf b})} \equiv H_{{\bf a}}H_{{\bf b}}$.  We could easily have
denoted it just as well by $H_{({\bf b},{\bf a})}=H_{{\bf b}}H_{{\bf a}}$.\\

On the contrary, the result in the case of the paired photon experiments under consideration
is that none of the four $H_{({\bf a}^*,{\bf b}^*)}$ matrices commute!  That is, for example,
$H_{({\bf a},{\bf b})}H_{({\bf a},{\bf b'})} \ \neq \ H_{({\bf a},{\bf b'})}H_{({\bf a},{\bf b})}$.  All this is to
say that the technical manipulations of mathematical quantum theory instantiate formally just what we knew to begin with ... that we cannot simultaneously perform the measurement observation of the polarization products at {\it both} angle settings $({\bf a},{\bf b})$ and  $({\bf a},{\bf b'})$ on the {\it same pair of photons}.\\

Well, who would want to?  We shall now find out.

\section{The principle of local realism and its relevance to Bell}

A feature crucial to the touted violation of Bell's inequality is that it
pertains to experimental results supposedly conducted with a {\it single photon pair} at all
four angle settings.  Sound unusual?  When the probabilistic pronouncements of quantum theory were formalized, Einstein
among others was puzzled by the fact that the conditional probability for the outcome of the experiment
at station $A$ depends on both the angle at which the experiment is conducted at station $B$ and on the
outcome of that experiment.  This matter is codified by the conditional probabilities we have seen in
equations (5).  This entanglement of seemingly unrelated physical processes was deemed to be a matter of
``spooky action at a distance''.  Well, Einstein proposed a solution to this enigma, positing that there
must be some other factors relevant to what might be happening at the
polarizer stations $A$ and $B$ that would account for the photon detections found to arise. As yet
unspecified in the theory, he considered such factors to identify unknown values of ``supplementary variables''.
 It was proposed that the probabilities inherent in the results of quantum theory must be
 representations of scientific uncertainty about the action
of these other variables on the two photons at their respective stations.  This was his way of accounting
for the spooky action at a distance.\\

However, there was one aspect of the matter upon which Einstein wanted to insist:  this was termed ``the
principle of local realism''.  Fair enough, quantum theory does stipulate the probability for
the photon detection at angle setting ${\bf a}$ as depending on whether the
polarizer direction at $B$ is set at ${\bf b}$ or ${\bf b'}$ and on what happens there.  However, in any specific
instance of the joint experiment at a relative polarization angle ${(\bf a},{\bf b})$, if the measurement observation
at $A$ happened to equal $A({\bf a},{\bf b}) = +1$, say, then in this instance the measurement at $A$ would have
to be the same no matter whether the direction setting at station B were ${\bf b}$ or ${\bf b'}$.  That is to
say, if the polarization observation $A({\bf a}) = +1$ in a particular experiment on a pair of photons
measure in the paired angle design $({\bf a},{\bf b})$, then the value of $A({\bf a})$ would also have to
equal $+1$ in a companion experiment on the same pair of photons when the polarization directions
would be set in the angle pairing $({\bf a},{\bf b'})$.\\

Actually, in a way we have already deferred to such an understanding.  We have been denoting the photon detection
value at station $A$ merely by $A({\bf a})$ rather than denoting it by $A({\bf a},{\bf b})$, even before we have now
introduced consideration of this principle of local realism.
In the context of locality, the importance of such simplification of the notation was stressed by Aspect (2004, p 13/34).
In fact there was no need to denote the paired direction
at $B$ in our notation earlier, because we can only do the experiment on a specific photon at one specific possibility
pair determining the angle pairing $({\bf a}^*,{\bf b}^*)$.  So we have merely denoted the measurements as
$A({\bf a})$ and $B({\bf b})$, or $B({\bf b'})$.  Nonetheless, the QM probabilities of equations $(1)$ stipulated
that each of the paired results of the experiments does depend jointly on the {\it relative} angle between the
two polarization directions.\\

Despite this notational deference, we should now recognize explicitly and declare loudly
that this principle of local realism is based upon a claim that lies  outside the bounds of matters addressed
by the theory of quantum physics.  For, as we have noted, it is impossible to make a measurement of both
the photon detection product $A({\bf a})B({\bf b})$ {\it and} the product $A({\bf a})B({\bf b'})$ {\it on the
same pair of photons.}  So quantum theory explicitly disavows addressing this matter directly.\\

We are ready to conclude this Section by proposing an experimental measurement that lies at the heart of
Bell's inequality.  We are not yet ready to assess it, nor to explain its relevance to the principle of local
realism, but we shall merely air it now for viewing.  Peculiar, it is considered to be the result of a
gedankenexperiment.\\

Consider a pair of photons to be ejected toward stations $A$ and $B$ at which the pair of
polarizers can be directed in any of the four relative angles we have described.  According to the detection of whether
the photons pass through the polarizers or are deflected by them, Bell's inequality pertains to an experimental
 quantity defined by the equation \vspace{.2cm}\\
\hspace*{1.5cm} $\ \ \ \ \ \ \ \ s \ \equiv \ A({\bf a})\,B( {\bf b})\; -\;
A({\bf a})\,B( {\bf b}')\; +\; A( {\bf a}')\,B( {\bf
b})\; +\; A( {\bf a}')\,B({\bf b}')\, \ \ . \vspace{.2cm}\hspace{2.3cm}(6)$\\
\noindent Mathematically, we would refer to this quantity $s$ as a linear combination of four polarization detection
products.  Any one of the four terms that determine the value of $s$ could be observed in an experiment on a pair of
prepared photons. Before we explain why this quantity is of interest, we should recognize right here only that
we {\it can} observe the value of this quantity $s$ if we are to conduct four component experiments on four distinct
pairs of photons, each ejected toward stations $A$ and $B$ with the polarizers directed at a different relative
angle pairing.  However, we {\it cannot} observe the value of $s$ if it were meant to pertain to all four experiments
being conducted on the same pair of photons.  It just cannot be done, and quantum theory is very explicit about
having nothing directly to say about its value. If we are to consider the value of $s$ in such an experimental
 design, it could only be as the result of a ``thought experiment''.  Enough said for now.\\

Why would we even be interested by such a ``gedankenexperiment'' as its perpetrators called it, and what does the supposition of ``hidden variables''
have to do with the matter?

\section {Einstein's proposal of hidden variables relevant to the matter}

Puzzled by the standing probabilistic conclusions of quantum theory which he helped to formulate, Einstein wondered
what could be the meaning of these ``probabilities'' involved in its prescriptions.  Others were proclaiming
that the experimental and theoretical discoveries of QM support the view that at its fundamental level of
particulate matter, the behaviour of Nature is random, and that quantum theory had identified its probabilistic
structure.  Convinced that ``he (the old one) does not play dice with the universe'', Einstein formulated another proposal:
that the
analysis of quantum theory is incomplete.  There must be some other matters involved in  quantum level
experimentation that we do not know about; and  these other unknown ``supplementary''
variables would conceivably be distinguishable in the observable outcome of any
particular experimental result if only we knew how to distinguish their measurable states.
The probabilities of quantum theoretical specifications must formalize our
uncertain knowledge of the situation, our uncertainty distribution concerning the conceivable
instantiation of these hidden
variables in any particular experimental setting.  \\

Enter the necessity to formulate a gedankenexperiment to assess the matter.  A paper by Einstein, Podolsky, and Rosen (1935)
presented this argument which became known subsequently as the EPR proposal. It stimulated a fury
of healthy discussion and argument that I shall not summarize here.  Well documented both in the professional journals of
physics and in literature of popular science, the discussion featured considerations of the collapse of a quantum system
when subject to observation that disturbs it, the non-locality of quantum processes, and esoteric formulations of the ``many worlds'' view
of quantum theory.  What matters for my presentation here is that Einstein's views were widely relegated as a
quirky peculiar
sideline, and the recognition of randomness as a fundamental feature of quantum activity came to the forefront of
theoretical physics.\\

Enter John Bell.  Interested in a reconsideration of Einstein's view, he began his research with an idea to
re-establish its validity as a contending interpretation of what we know.  However, he was surprised to find
this programme at an impasse when he discovered that the probabilistic specifications of quantum theory which
we have described above seem to defy a simple requirement of mathematical
probabilities, if the principle of local realism is valid.  In the context of a hidden variables interpretation
of the matter, this seemed to require that the principle of local realism must be rejected.
 Reported in a pair of articles (Bell, 1964, 1966), these results too stimulated a continuation of the
flurry which has lasted through the 2015 publication in {\it Nature} of their apparently definitive substantiation by the research at
the Delft University of Technology.   \\

The specification of Bell's inequality can take many forms.  The context in which it is addressed in the
remainder of my exposition here was presented in an article by Clauser, Horne, Shimony, and Holt (1969), commonly
referred to as the CHSH formulation.  This was the form that attracted still another principal investigator in
this story, Alain Aspect.  A young experimentalist, he wondered how could such a monumental result of quantum physics
pertain only to a thought experiment, devoid of actual physical experimental confirmation.  He thought to have
devised an experimental method that could confirm or deny the defiance of Bell's inequality.  My assessment of
his empirical work follows directly from his explanation of the situation (Aspect, 2002) reported to a conference organized to
memorialize Bell's work.  My notation is largely the same as Aspect's.  I adjust only the notation for expectation
of a random variable to the standard form of $E(X)$, replacing his notation of $<X>$ which has become standard in
mathematical physics in the context of bra-ket notation which I avoid.  Here is how it works.

\subsection{Explicit construction of $s$ with hidden variables}

Hidden variables theory proposes that the quantity $s$ which we have introduced in equation $(6)$ should be considered
to derive from a physical function of unobserved and unknown hidden variables, whose values might be codified
by the vector $\lambda$, viz., \vspace{.2cm} \hspace{5.5cm}\\
$\hspace*{1cm} s(\lambda) \ \equiv \ A(\lambda, {\bf a})\,B(\lambda, {\bf b})\; -\;
A(\lambda, {\bf a})\,B(\lambda, {\bf b}')\; +\; A(\lambda, {\bf a}')\,B(\lambda, {\bf
b})\; +\; A(\lambda, {\bf a}')\,B(\lambda, {\bf b}')\, , \ $ \vspace{.2cm} \ \ (7)\\
\noindent for $\lambda \ \in
\Lambda$\,.  The variable designated by $\lambda$ here could be a vector of any number of
components identifying unknown features of the experimental setup that are relevant to the outcome of the
experiment in any specific instantiation. The set designated by $\Lambda$ is meant to represent the space of
possible values of theses hidden variables.  The status of these variables in the context of any particular
experiment is supposed only to depend on the state of the photon pair and its surrounds, independent of the angle setting $({\bf a}^*,{\bf b}^*)$ at which the polarizers are directed.  According to the deterministic outlook underlying the physical theory relying on recourse to the hidden
variables, if we could only know the values of these hidden variables at the time of any experimental run
and have a complete theoretical understanding of
their relevance to the polarization behaviour of the photon pair, then we would know what would be the values of the
polarization incidence detection of the photon pair at any one or all of possible angle settings.\\

Now the personalist subjective theory of probability (apparently subscribed to by Einstein, and surely by
Bruno de Finetti and by me)  specifies that any individual's uncertain knowledge of the values of
observable but unknown quantities could be representable by a probability density
over its space of possibilities.   Aspect denotes such a density in this situation by $\rho(\lambda)$.
For any proponent of the quantum probabilities it might well be presumed to be ``rotationally
invariant'' over the full $360^{\circ}$ of angles at which the photon may be fluttering toward the polarizer.
That is
to say, the probabilities for the possible values of the supplementary variables do not depend on the angular
direction in $(x,y)$ dimensions of the photons along ${\bf z}$ axes heading toward stations $A$ and $B$.\\

Since we avowedly have no idea of what these hidden variables  might be, much less what their
numerical values may be relevant to any specific experimental run, we can only ponder the ``expected value'' of
$s(\lambda)$ with respect to the distribution specified by $\rho(\lambda)$.
The feature of rotational invariance implies that this expectation
is the same no matter what be the rotational angle at which the photons flutter relative to their $(x,y)$ plane detections.  Let's write this expectation equation down:\vspace{.2cm}\\
$E[s(\lambda)] \, \vspace{.3cm} = \,E[A(\lambda, {\bf a})B(\lambda, {\bf b})] -
E[A(\lambda, {\bf a})B(\lambda, {\bf b}')] + E[A(\lambda, {\bf a}')B(\lambda, {\bf b})] +
E[A(\lambda, {\bf a}')B(\lambda, {\bf b}')]. \ \ \ \ (8)$\vspace{.2cm}\\
\noindent This equation follows directly from equation $(7)$ because of the fact that a rule of
probability says that the expectation of any linear combination of random quantities equals the same
linear combination of their expectations.  Fair enough.  Now fortunately, we have already reported in equation $(5)$
that the probabilities of quantum theory identify the expected value of any polarization product at the variable
relative polarization angle $({\bf a}^*,{\bf b}^*)$ as  $E[A({\bf a}^*)B({\bf b}^*)] = cos\ 2 ({\bf a}^*,{\bf b}^*)$.
So we are ready to proceed.

\subsection{Finally, Bell's inequality}

We have now arrived at a place we can state precisely what Bell's inequality says. There is just
a little more specificity to detail before we soon will have it.  However, I should alert you that there is a little
tic in the understanding of equation $(8)$ to which we shall return after we learn how the inequality is currently understood
to be defied by quantum theory.  But on the face of it, the validity of equation $(8)$ is plain as day.\\

Now re-examining equation $(7)$, it is apparent that it can be factored into a simplified form:\vspace{.2cm}\\
$\ \ \ \ \ \ \ \ s(\lambda) \ \equiv \ A(\lambda, {\bf a})\,B(\lambda, {\bf b})\; -\;
A(\lambda, {\bf a})\,B(\lambda, {\bf b}')\; +\; A(\lambda, {\bf a}')\,B(\lambda, {\bf
b})\; +\; A(\lambda, {\bf a}')\,B(\lambda, {\bf b}')\, , \ $ for $\lambda \ \in
\Lambda$\,,\vspace{.2cm}\\
$\hspace*{1cm} = \ A(\lambda, {\bf a})\; [B(\lambda, {\bf b}) - B(\lambda, {\bf b}')]\ +
\ A(\lambda, {\bf a}')\; [B(\lambda, {\bf b}) + B(\lambda, {\bf b}')]
\vspace{.2 cm}$\ , \ \ \ \ \  and alternatively \ \\
$\hspace*{1cm}= \ B(\lambda, {\bf b})\; [A(\lambda, {\bf a}) + A(\lambda, {\bf a}')]\ - \
B(\lambda, {\bf b}')\; [A(\lambda, {\bf a}) - A(\lambda, {\bf a}')] \ .\ $  \ \ \ \ \ \hspace{3cm}$(9)$
\vspace{.3cm}\\
\noindent It is important to notice that once again, in performing this simple factorization of the components
$A(\lambda, {\bf a})$ and $A(\lambda, {\bf a'})$ in this second line, we have implicitly presumed the principle
of local realism.  For when we consider the first two summands of the first line,
$A(\lambda, {\bf a})\,B(\lambda, {\bf b})$, and $A(\lambda, {\bf a})\,B(\lambda, {\bf b}')$, we should notice
that the value of $A(\lambda, {\bf a})$ in that first term is evaluated in an experiment at which the paired
polarization angle is $({\bf a},{\bf b})$, whereas in the second term from which it is factored it is evaluated in
an experiment at the relative polarization angle $({\bf a},{\bf b'})$.  It is the principle of local realism,
extraneous to any claims of quantum theory, that provides the observed value of $A(\lambda,{\bf a})$
must be identical
in these two conditions which are impossible to instantiate together. It is only under the condition of this
assertion that we would be able to factor this term out of the two expressions. The same goes for the factorization
 of $A(\lambda, {\bf a'})$.  This is not a source of any
worry.  I am merely mentioning this so that we are all aware of what is going on.  The same feature of supposition
is pertinent to the alternative factorization of the terms $B(\lambda, {\bf b})$ and  $B(\lambda,{\bf b'})$ in the
third line
from the terms of the first line.\\

Having arrived at  this factorization, it will now take just a little thought to recognize that if the value
of the quantity $s$ is supposed to be determined from a thought experiment on a single pair of photons, then the
numerical value of $s$ can equal only either $+2$ or $-2$.  Of course, if
we were to calculate the value of $s$ from
performing four component experiments with {\it four different pairs} of photons (something we can actually do), then the four component product
values might each then equal either $-1$ or $+1$, so the
value of $s$ might equal any of $\{-4, -2, 0,  +2, +4\}$. However, in such a case the factorization we performed in equation $(9)$ would
not be permitted.
For each of the observed detection products appearing in the first line would pertain to
a different pair of photons whose multiplicands would be free to equal either $+1$ or $-1$ as prescribed by experiment.
The same possibilities would be accessible if the principle of local
realism were not valid.  However, if the value of $s$ is to be calculated from the results
of a thought experiment on the same pair of photons, then its possibilities would be limited according to local realism merely to
$\{-2, +2\}$. Here is how to recognize this.  \\

 Suppose the values of
$B(\lambda, {\bf b})$ and $B(\lambda, {\bf b}')$ were both observed to equal $+1$.  Then the first term in the factored
form of the second line
must equal $A(\lambda, {\bf a})\; [B(\lambda, {\bf b}) - B(\lambda, {\bf b}')] \ = \ 0$;  and furthermore, the second term in the factored representation would then be $A(\lambda, {\bf a'})\; [B(\lambda, {\bf b}) + B(\lambda, {\bf b}')]$.
The factor $A(\lambda, {\bf a'})$ equaling either $+1$ or $-1$ would then be multiplied by the factor
$[B(\lambda, {\bf b}) + B(\lambda, {\bf b}')]$ which would equal
the number $+2$.  Thus, the value of $s$ could equal only either $-2$ or $+2$.
Alternatively, suppose that the values of $B(\lambda, {\bf b})$ and $B(\lambda, {\bf b}')$ are both observed to equal $-1$.  Then by a similar argument the value of the first factored expression would again equal $0$ and the second expression
would equal either $-1$ or $+1$ multiplied now by $-2$.  Again the computed result of the value of $s$ could equal
only $-2$ or $+2$.  I leave it to the reader to confirm the same result for the possible values of $s$ if the values
of $B(\lambda, {\bf b})$ and $B(\lambda, {\bf b}')$ were observed to equal either
$-1$ and $+1$ respectively, or $+1$ and $-1$ respectively.\\

The conclusion is indisputable.  If the principle of local realism holds, then the value of $s$
that would be instantiated as a result of a thought experiment on the same pair of photons in all four polarization
angle settings can equal only $-2$ or $+2$.  Thus, the expected value $E(s)$ deriving from {\it any} coherent probability distribution over the
four values of the component paired polarization experiments would have to be a number between $-2$ and $+2$.  Stated algebraically and simply,
without all the provisos explaining its content, Bell's inequality is the requirement that $-2 \leq E(s) \leq +2$.\\

Well, what do the probabilities of quantum theory imply for the value of $E(s)$\ ?  The answer universally presumed to
be correct by proponents of the Bell violation is that when the design of the four experiments on a single pair of
 photons is constructed at a particular array of angle settings that we shall soon identify, then $E(s) \ = \ 2\sqrt{2} \ = \ 2.8284$ to four decimal places,
a number that exceeds $+2$. (I shall show you why in the next paragraph.)  Moreover, the experimental results of Aspect, as well as the more
sophisticated experimentation of succeeding decades, is understood to corroborate this result to many decimal places.
I will soon explain how this result is derived as well.  However, I will insist on also showing you that not only is this theoretical
derivation wrong, but that the calculations used to corroborate this result from experimental evidence are mistaken.
Nonetheless, there is nothing at all wrong with the experimental results, which are what they are.


\subsection{The mistaken violation of Bell's inequality}

It turns out that Bell's inequality is {\it not} deemed to be
defied at {\it every} four-plex of possible experimental angle
settings that we have characterised generically as  $({\bf a},{\bf b})$,  $({\bf a},{\bf b'})$,  $({\bf a'},{\bf b})$, or \ $({\bf a'},{\bf b'})$.
At some paired directional settings of the polarizers it seems not to be defied at all.  Among other
pairings at which it seems to be defied, it is apparently defied more strongly at some pairings than at others.  Aspect had thought that
if we were to find experimental evidence of the defiance, we should try to find it at the angle pairings for which the
theoretical defiance is the most extreme.  It is a matter of simple calculus of extreme values to discover that the
most extreme violation of the equality should occur at the angle settings $({\bf a},{\bf b}) = -\pi/8$,  $({\bf a},{\bf b'}) = -3\pi/8$,  $({\bf a'},{\bf b}) = \pi/8$, or \ $({\bf a'},{\bf b'}) = -\pi/8$.
(The angle measurements are expressed here in terms of their polar representations.  In terms of degrees, the angle $-\pi/8  = -22.5^{\circ}$,
while $-3\pi/8 = -67.5^{\circ}$, and $+\pi/8  = +22.5^{\circ}$.)  You may wish to examine our
Figure 2 and notice that the angles between the various polarization directions we depicted there correspond to
these relative angles.  For the record, doubling these angles  yields the values of
$\pm\pi/4  = \pm 45^{\circ}$ and $-3\pi/4 = -135^{\circ}$ in these instances.  And why does that matter? ... \\

Recall equation $(8)$ and the ensuing sentences.  Evaluating $E(s)$ according to this equation at the four angle
settings just mentioned requires evaluating the summand component expectations.  Each of them in the
form $E[A({\bf a}^*)B({\bf b}^*)] = cos\ 2({\bf a}^*,{\bf b}^*)$, these would then be 
\begin{alignat*}{5}
E[A(\lambda, {\bf a})B(\lambda, {\bf b})]\  &=\  cos\ 2({\bf a},{\bf b}) &&= \ cos(-\pi/4) &&= \ \ 1/\sqrt{2}\ ,\nonumber \vspace{.2cm}\\
E[A(\lambda, {\bf a})B(\lambda, {\bf b'})]\  &=\  cos\ 2({\bf a},{\bf b'}) &&= \ cos(-3\pi/4) &&= -1/\sqrt{2}\ ,\nonumber \vspace{.2cm}\\
E[A(\lambda, {\bf a'})B(\lambda, {\bf b})]\  &=\  cos\ 2({\bf a'},{\bf b}) &&= \ cos(\pi/4) &&= \ \ 1/\sqrt{2}\ , \ \ {\rm and}\nonumber \vspace{.2cm}\\
E[A(\lambda, {\bf a'})B(\lambda, {\bf b'})]\  &=\  cos\ 2({\bf a'},{\bf b'}) &&= \ cos(-\pi/4) &&= \ \ 1/\sqrt{2}\ ,\nonumber \vspace{.3cm}
\end{alignat*}
\hspace*{.1cm}apparently yielding \ \ $E[s(\lambda)] \ \ \vspace{.3cm} = \ \ 1/\sqrt{2} \ - \ (-1/\sqrt{2}) \ + \ 1/\sqrt{2} \ + \ 1/\sqrt{2} \ \ = \ \ 4/\sqrt{2} \ = \ \ 2\sqrt{2}\ .$\\
\noindent Voila!  The expected value of $s$ apparently equals $2\sqrt{2} \approx 2.8284$, \ a real number outside of the interval  $[-2,+2]$,
defying Bell's inequality!  What could be more simple, direct, and stunning? \\

Answer: . . . the truth! \ \ \ . . . OK, what is wrong, if anything?\\

The answer is seen most simply by constructing and then examining a matrix, which in the jargon of the operational
subjective theory of probability is called ``the realm matrix of possible observation values''
that could result from the performance of the gedankenexperiment in CHSH form.
I will display this entire matrix on the next page, in a partitioned
form of its full extension as it pertains
to every aspect of the problem we shall discuss. Then we shall discuss it, piece by piece.  I should mention here that
while the name ``realm matrix of possibilities'' has arisen from within the operational subjective construction of the
theory of probability, the matrix itself is merely a well-defined matrix of numbers that can be understood and
appreciated by any experimentalist, no matter what may be your personal views about the foundations of probability.
In the jargon of quantum physics it might be called the ensemble matrix of possible observation vectors.

\section{A neglected functional dependence}

In specifying the QM motivated expectation $E[s(\lambda)]$ as they do in our equation $(8)$, Aspect/Bell fail to
recognize a {\it symmetric functional dependence} among the values of the four proposed polarization
products composing $s(\lambda)$ as defined in equation $(7)$,
when it is meant to correspond to the result of the 4-ply thought-experiment on the same pair of photons.  Perhaps
surprisingly, the achieved values of any three {\it products} of the paired polarization
indicators imply a unique value for the fourth product.  We now engage to substantiate this claim.

\subsection{The realm matrix of experimental quantities}

Consider the realm matrix of all quantities relevant to the observations that might be made in the
proposed 4-ply gedankenexperiment on a pair of photons under investigation.  
On the left side of the realm equation is written
the name ${\bf R}({\bf X})$, where ${\bf X}$ is a partitioned vector of names of every quantity that will
be relevant to the outcome of the experiment and what quantum theory asserts about it. 
You will already recognize those in the first two partitioned blocks.
On the right side of
the realm equation appears a matrix whose columns exhaustively identify the values of these partitioned quantities
that could possibly result from conducting the gedankenexperiment.  We shall discuss them in turn.\\

\noindent \hspace*{1.5cm}${\bf R}\left(\begin{array}{c}
A({\bf a}) \\
B({\bf b}) \\
A({\bf a}')\\
B({\bf b}') \\
***** \\
A({\bf a})B({\bf b}) \\
A({\bf a})B({\bf b}') \\
A({\bf a}')B({\bf b}) \\
A({\bf a}')B({\bf b}') \\
***** \\
\mathcal{A}({\bf a}')\mathcal{B}({\bf b}') \\
***** \\
\Sigma_{/({\bf a},{\bf b})}\\
\Sigma_{/({\bf a},{\bf b}')}\\
\Sigma_{/({\bf a}',{\bf b})}\\
\Sigma_{/({\bf a}',{\bf b}')}\\
*****\\
s(\lambda)\\
s_{\mathcal{A/B}({\bf a}',{\bf b}')}\\
1
%
%
\end{array}\right) \ \ = \ \
\left(\begin{array}{cccccccccccccccc}
1 & 1 & 1 & 1 & \llap{$-$}1 & \llap{$-$}1 & \llap{$-$}1 & \llap{$-$}1 & 1 & 1 & 1 & 1 &
\llap{$-$}1 & \llap{$-$}1 & \llap{$-$}1 & \llap{$-$}1 \\
1 & 1 & 1 & 1 & 1 & 1 & 1 & 1 & \llap{$-$}1 & \llap{$-$}1 & \llap{$-$}1 & \llap{$-$}1 &
\llap{$-$}1 & \llap{$-$}1 & \llap{$-$}1 & \llap{$-$}1 \\
1 & 1 & \llap{$-$}1 & \llap{$-$}1 & 1 & 1 & \llap{$-$}1 & \llap{$-$}1 & 1 & 1 &
\llap{$-$}1 & \llap{$-$}1 & 1 & 1 & \llap{$-$}1 & \llap{$-$}1 \\
1 & \llap{$-$}1 & 1 & \llap{$-$}1 & 1 & \llap{$-$}1 & 1 & \llap{$-$}1 & 1 & \llap{$-$}1 &
1 & \llap{$-$}1 & 1 & \llap{$-$}1 & 1 & \llap{$-$}1 \\
& & & & & & & & & & & & & & & \\
1 & 1 & 1 & 1 & \llap{$-$}1 & \llap{$-$}1 & \llap{$-$}1 & \llap{$-$}1 & \llap{$-$}1 &
\llap{$-$}1 & \llap{$-$}1 & \llap{$-$}1 & 1 & 1 & 1 & 1 \\
1 & \llap{$-$}1 & 1 & \llap{$-$}1 & \llap{$-$}1 & 1 & \llap{$-$}1 & 1 & 1 & \llap{$-$}1 &
1 & \llap{$-$}1 & \llap{$-$}1 & 1 & \llap{$-$}1 & 1 \\
1 & 1 & \llap{$-$}1 & \llap{$-$}1 & 1 & 1 & \llap{$-$}1 & \llap{$-$}1 & \llap{$-$}1 &
\llap{$-$}1 & 1 & 1 & \llap{$-$}1 & \llap{$-$}1 & 1 & 1 \\
1 & \llap{$-$}1 & \llap{$-$}1 & 1 & 1 & \llap{$-$}1 & \llap{$-$}1 & 1 & 1 & \llap{$-$}1 &
\llap{$-$}1 & 1 & 1 & \llap{$-$}1 & \llap{$-$}1 & 1 \\
& & & & & & & & & & & & & & & \\
1 & 1 & 1 & 1 &1 & 1 & 1 & 1 & \llap{$-$}1 & \llap{$-$}1 & \llap{$-$}1 & \llap{$-$}1 &
\llap{$-$}1 & \llap{$-$}1 & \llap{$-$}1 & \llap{$-$}1 \\
& & & & & & & & & & & & & & & \\
3 & \llap{$-$}1 & \llap{$-$}1 & \llap{$-$}1 & 1 & 1 & \llap{$-$}3 & 1 & 1 & \llap{$-$}3
& 1 & 1 & \llap{$-$}1 & \llap{$-$}1 & \llap{$-$}1 & 3 \\
3 & 1 & \llap{$-$}1 & 1 & 1 & \llap{$-$}1 & \llap{$-$}3 & \llap{$-$}1 & \llap{$-$}1 &
\llap{$-$}3 & \llap{$-$}1 & 1 & 1 & \llap{$-$}1 & 1 & 3 \\
3 & \llap{$-$}1 & 1 & 1 &\llap{$-$}1 & \llap{$-$}1 & \llap{$-$}3 & 1 & 1 & \llap{$-$}3 &
\llap{$-$}1 & \llap{$-$}1 & 1 & 1 & \llap{$-$}1 & 3 \\
3 & 1 & 1 & \llap{$-$}1 & \llap{$-$}1 & 1 & \llap{$-$}3 & \llap{$-$}1 & \llap{$-$}1 &
\llap{$-$}3 & 1 & \llap{$-$}1 & \llap{$-$}1 & 1 & 1 & 3 \\
& & & & & & & & & & & & & & & \\
2 & 2 & \llap{$-$}2 & 2 & 2 & \llap{$-$}2 & \llap{$-$}2 & \llap{$-$}2 & \llap{$-$}2 &
\llap{$-$}2 & \llap{$-$}2 & 2 & 2 & \llap{$-$}2 & 2 & 2 \\
2 & 4 & 0 & 2 & 2 & 0 & 0 & \llap{$-$}2 & \llap{$-$}4 & \llap{$-$}2 & \llap{$-$}2 & 0 & 0
& \llap{$-$}2 & 2 & 0\\
1 & 1 & 1 & 1 & 1 & 1 & 1 & 1 & 1 & 1 & 1 & 1 & 1 & 1 & 1 & 1
\end{array}\right)$ \vspace{.3cm}

The sixteen columns of four-dimensional vectors in the first partitioned block exhaustively list all
the speculative $4 \times 1$ vectors of observation values that could possibly arise among the four
experimental detections of photons
at the four angles of polarizer pairings.  In order to observe the detection {\it products} at the four relative angles
$ A({\bf a})B({\bf b}), \, A({\bf a})B({\bf b}'), \, A({\bf a}')B({\bf b}),\,$
and $A({\bf a}')B({\bf b}')\,$, we would surely have to observe each of the four {\it multiplicands} involved in their
specification: \ $A({\bf a}), \, B({\bf b}), \, A({\bf a'})$, and $B({\bf b'})$.  
Since each of these observation
values might equal only either $-1$ or $+1$, there are sixteen possibilities of the 4-dimensional result of the
4-ply experiment.  There are no presumptions made about these prospective quantity values: neither
whether they ``exist'' or
not prior to the conduct of the experiment at all, nor even whether they exist in any form after the experiment is
conducted.  We have merely made a list of what we could possibly observe if indeed we were capable of conducting
the proposed gedankenexperiment on the same pair of photons.  The observation vector would have to equal one of the
16 columns appearing in the top bank of the partitioned realm matrix.\\


Every other component quantity in the columns displayed in subsequent blocks of the realm matrix is computed via some function of these possibilities. Notice once again that the ``exhaustiveness'' of this list presupposes the principle of local realism, specifying for example that the value of $A({\bf a})$ identifying whether the photon passes through the polarizer at $A$ or not, would be the same no matter whether the polarizer at which the paired photon engages station $B$ is set at direction ${\bf b}$ or at${\bf b'}.$\\

To begin the completion of the realm matrix, the second block of components identifies the four
designated {\it products} of the paired polarization indicators that yield the value of the quantity
$s$ as it is simply defined in equation $(6)$. The first row of this second block, identifying the
product $ A({\bf a})B({\bf b})$, is the componentwise product of the first two rows of
the first block.  The second row of this block, identifying the product $ A({\bf
a})B({\bf b}')$, is the componentwise product of the first and fourth rows of the first
block, and so on.  This second block lists exhaustively all the combinations of polarization products
that we could possibly observe in the conduct of our gedankenexperiment.  Examine any one of
these columns of products, checking that in fact the value of each product in that column is
equal to the product of the corresponding multiplicands appearing in the column directly above it.\\

The first item to notice about this realm matrix is that, whereas the sixteen columns of
the first block of polarization observations are distinct, the second block contains only
{\it eight} distinct column vectors.  Columns 9 through 16 in block two of the realm
matrix reproduce columns 1 through 8 in reverse order.  Moreover, examining the first
{\it three} rows of this second block more closely, it can be recognized that the first
eight columns of these rows exclusively exhaust the simultaneous measurement
possibilities for the three product quantities  they identify.  These are the eight
vectors of the cartesian product $\{+1, -1\}^3\,$, which are repeated in columns nine
through sixteen in reverse order.  Together, what these two observations mean is that the fourth product
quantity in this second block of vector components is derivable as a function of the
first three.  What is more, any one of the product quantities identified in block two is
determined by the same computational function of the other three!  This is what I meant
earlier when alluding that the photon detection products in the gedankenexperiment have
embedded within them four symmetric functional relations.  This can be seen by
examining the columns of the {\it fourth} block of the matrix, which we shall do
shortly.\\

The third block of the realm matrix contains only a single row, corresponding to a quantity
we designate as $\mathcal{A}({\bf a}')\mathcal{B}({\bf b}')$.  This quantity takes
values only of $\pm1$, but it is logically independent of the product quantities
appearing in the first three rows of block two.  This is the quantity that Aspect/Bell
think they are assessing when they freely specify the quantum expectations for all four
angle settings as they do, seemingly defying Bell's inequality.  We denote its name
with calligraphic type to distinguish it from the actual polarization product $A({\bf
a}')B({\bf b}')$ whose functional relation to the other three products we are now
identifying. Peculiar, this singular component of the fourth partition block is not an
``Alice and Bob'' observation quantity, but rather an
``Aspect/Bell'' imagined quantity.  It is logically independent of the first
three ``Alice and Bob'' products.  This is to say that whatever values these products may be, the value of
$\mathcal{A}({\bf a}')\mathcal{B}({\bf b}')$ may equal $+1$ in the appropriate row
among the first eight columns, or it may equal $-1$ in the corresponding column among
the second eight.  However, it does {\it not} represent the photon detection product $A({\bf
a}')B({\bf b}')$ in the four imagined experiments on a single photon pair.

\subsection{Specifying the functional form via block four}

Quantities in the fourth block of the realm matrix are designated with the names
$\Sigma_{/({\bf a},{\bf b})}, \Sigma_{/({\bf a},{\bf b}')}$,  $\Sigma_{/({\bf a}',{\bf
b})}$, and $\Sigma_{/({\bf a}',{\bf b}')}$.  These quantities are defined by sums of
column elements in {\it those rows of the second block} that are {\it not marked} behind
the slash in the notational subscript.  For examples, \\
 \hspace*{3.1cm}$\Sigma_{/({\bf a},{\bf b})}\ \equiv \ A({\bf a})B({\bf b}') \; + \;
A({\bf a}')B({\bf b}) \; + \; A({\bf a}')B({\bf b}')$, \ \ \ \ \ \ and \vspace{.2cm}\\
 \hspace*{3cm}$\Sigma_{/({\bf a},{\bf b}')} \ \equiv  \   \,A({\bf a})B({\bf b})\, \; + \;
    A({\bf a}')B({\bf b}) \; + \;
    A({\bf a}')B({\bf b}')\ . $ \vspace{.2cm}\\
 \noindent The quantities $\Sigma_{/({\bf a}',{\bf b})}$, and $\Sigma_{/({\bf a}',{\bf
b}')}$ are defined similarly.\\

Next to notice is that the fourth row of the {\it second} matrix block, corresponding to
$A({\bf a}')B({\bf b}')$, has an entry of $1$ if and only if the fourth row of the {\it
fourth} block, corresponding to $\Sigma_{/({\bf a}',{\bf b}')}$, has an entry of $-1$ or
$+3$ in the same column.  When that entry is $+1$ or $-3$, the corresponding entry of the
second block is $-1$.  What this recognition does is to identify the functional relation
of the fourth polarization product to the first three polarization products, viz.,
\vspace{.1cm} \\
\noindent\hspace*{2.8cm} $A({\bf a}')B({\bf b}') \ = \ {\bf G}[A({\bf a})B({\bf b}),\,A({\bf a}')B({\bf b}),\,
    A({\bf a})B({\bf b}')] \\
\hspace*{4.75cm} \ \equiv \
\left(\Sigma_{/({\bf a}',{\bf b}')} = -1
\; or \; +3\right) \ - \ \left(\Sigma_{/({\bf a}',{\bf b}')} = +1 \; or \, -3\right) \  $.\hspace{1.5cm}$(10)$\\
\noindent Here and throughout this note I am using notation in which parentheses
surrounding a mathematical statement that might be true and might be false signifies the
number $1$ when the interior statement is true, and signifies $0$ when it is false.\\

Some eyeball work is required to recognize functional relationship $(10)$ by examining the
final row of block two and of block four together.  It may take even more concentration to recognize
that this very same functional rule identifies each of the
other three polarization products as a function of the other three as well!  The four
product quantities $A(^.)B(^.)$ are related by four symmetric functional relationships,
each of them being calculable via the same functional rule
applied to the other three!  This surprising recognition identifies the
source of the Aspect/Bell error in assessing the QM-motivated expectation for
$s(\lambda)$ in the way they do.\\

It is surely true that $E[s(\lambda)]$ equals a linear combination of four expectations of polarization products, as specified in equation $(8)$.
Moreover, if the definition of $s(\lambda)$ in equation $(6)$ were understood to represent the combination of observed products from experiments
on four distinct pairs of photons, then the possible values of $s(\lambda)$ would span the integers  $\{-4, -2, 0, 2, 4\}$;  the expectation of each product $E[A({\bf a}^*)B({\bf b}^*)]$ would equal $-1/\sqrt{2}$ or $+1/\sqrt{2}$ as appropriate to the angle $({\bf a}^*,{\bf b}^*)$;  and $E[s(\lambda)]$ would equal $2\sqrt{2}$ as proposed by Aspect/Bell. This involves no violation of any probabilistic inequality at all, and there is no suggestion of mysterious activity of quantum mechanics.\\

However, when it is proposed that the paired polarization experiments at all four considered angles pertain to the same photon pair, then each of the products is restricted to equal the specified function value of the other three that we identified explicitly for $A({\bf a}')B({\bf b}')$
in equation $(10)$ as $\Sigma_{/({\bf a}',{\bf b}')}$ via the function ${\bf G}[A({\bf a})B({\bf b}),\,A({\bf a}')B({\bf b}),\,
    A({\bf a})B({\bf b}')]$. In this context, Aspect's expected quantity would be representable equivalently by any
of the following equations:\vspace{.3cm}\ \ \ 
\noindent$E[s(\lambda)]  \vspace{.3cm} = E[A({\bf a})B({\bf b})] - E[A({\bf a})B({\bf b}')] + E[A({\bf a}')B({\bf b})] +
E\{{\bf G}[A({\bf a})B({\bf b}),A({\bf a})B({\bf b}'),
    A({\bf a}')B({\bf b})]\}$ \\
$\hspace*{.7cm} \vspace{.3cm} = \,E[A({\bf a})B({\bf b})] - E[A({\bf a})B({\bf b}')] + E[A({\bf a}')B({\bf b}')] +
E\{{\bf G}[A({\bf a})B({\bf b}),A({\bf a})B({\bf b}'),
    A({\bf a}')B({\bf b}')]\}$\\
$\hspace*{.65cm} \vspace{.3cm} = \,E[A({\bf a})B({\bf b})] + E[A({\bf a}')B({\bf b})] + E[A({\bf a}')B({\bf b}')] -
E\{{\bf G}[A({\bf a})B({\bf b}),A({\bf a}')B({\bf b}),
    A({\bf a}')B({\bf b}')]\}$ \\
$\hspace*{.6cm} \vspace{.3cm} = \,-E[A({\bf a})B({\bf b}')] + E[A({\bf a}')B({\bf b})] + E[A({\bf a}')B({\bf b}')] +
E\{{\bf G}[A({\bf a})B({\bf b}'),A({\bf a}')B({\bf b}),
    A({\bf a}')B({\bf b}')]\}$.\\
\hspace*{15.5cm}\vspace{.2cm}(11)\\
\noindent The symmetries imposed on this problem would yield an identical result in each case, which would surely
{\it not} yield $2\sqrt{2}$ at all.  {\it This is the mathematical error of neglect to which the title of this
current exposition alludes.}  What might it yield?\\

The functional relation we have exposed in $(10)$ is {\it not} linear.  If it were, then
the specification of an expectation for its arguments would imply the expectation value
for the function value.  As it is not, the specification of expectation values for the
arguments only imply {\it bounds} on any cohering expectation value for the fourth.  These
numerical bounds can be computed using a theorem due to Bruno de Finetti which he first
presented at his famous lectures at the Institute Henri Poincar\'e in 1935. He named it
only in his swansong text (de Finetti, 1974).  It was first characterized in the form of a
linear programming problem by Bruno and Giglio (1980), and has appeared in various forms
in recent decades. Among them are presentations in dual form by Whittle (1970, 1971) using
standard formalist notation and objectivist concepts.  We shall review the content of de
Finetti's theorem shortly, and then examine its relevance to assessing the expectation of
$s(\lambda)$ motivated by considerations of quantum mechanics.  We need first to air some
further brief remarks about the final block of the realm matrix.

\subsection{The remaining block of quantities and their realm components}

The first row of block five of the realm matrix merely identifies the values of
$s(\lambda)$ associated with the polarization observation possibilities enumerated in the
columns of block one.  Each component of this row is computed from the corresponding
column of block two according to the defining equation $(1)$. It is evident that every entry
of this row is either  $-2$ or $+2$.  This corresponds to the argument we have made
following the factorization equation $(9)$ in Section 5.
 The second row of this
block pertains to a quantity denoted as $s_{\mathcal{A/B}({\bf a}',{\bf b}')}$.  Its
value is defined similarly to equation $(6)$, but its final summand is specified as the
Aspect/Bell quantity $\mathcal{A}({\bf a}')\mathcal{B}({\bf b}')$ rather than the actual
polarization product quantity $A({\bf a}')B({\bf b}')$ that appears in this equation
defining $s(\lambda)$.  Again peculiar, its realm can be seen to include the elements
$\{-4,-2,0,2,4\}$ whereas the realm of $s(\lambda)$ includes only $\{-2,2\}$.  The fact
that the possibilities for $s_{\mathcal{A/B}({\bf a}',{\bf b}')}$ include both $-4$ and $+4$ is what makes
it not surprising that the expectation of {\it this quantity} is $2\sqrt{2}$ as
pronounced by proponents of the Aspect/Bell analysis.\\

The third row of block five is merely an accounting device, denoting that the ``sure''
quantity, $1$, is equal to $1$ no matter what  the observed results of the four
imagined optic experiments of Aspect/Bell might be.  Its relevance will become apparent when the
need arises to apply de Finetti's fundamental theorem to quantum assertions.\\

It is time for a rest and an interlude. It is a mathematical interlude whose complete understanding
 relies only on your knowledge of some basic methods of linear algebra.  If you would like a slow
 didactic introduction to the subject, my best suggestion is to look at Chapter 2.10 of my book,
 Lad (1996).  You may even wish to start in Section 2.7.  Another purely computational
 presentation appears in the article of Capotorti et al (2007, Section 4).  I will make another
 attempt here in a brief format, merely to keep this current exposition self-contained.
 What does the fundamental theorem of probability say?

\section{The relevance of the fundamental theorem of probability}

In brief, the fundamental theorem says that if you can specify expectation values
for any vector of quantities whatsoever, then the rules of  probability
provide  numerical bounds on a cohering expectation for any other quantity
 you would like to assess.  These can be computed from
the compilation of a linear programming routine. If the expectations you have specified
 are incoherent (meaning self-contradictory) among themselves,
then the linear programming problems they motivate have no
 solution.  This theorem is immediately relevant to our
situation here in which we have identified quantum-theory-motivated expectations
for any three of the four detection products that determine the value of $s$ for the
gedankenexperiment.  We wish to find the bounds on the cohering expectation for the
fourth detection product which is restricted to equal a function value determined by these three.  A discursive pedagogical
introduction is available in Lad (1996, 3.10).  In brief, here is how the
theorem works.\\


Suppose you have identified the expectations for $N$ quantities, and you are wondering
what you might assert as the expectation for another one, call it the $(N+1)^{st}$.  What
you should do to assess your sensible possibilities is firstly to
construct the realm matrix of possible values for the vector
of all $(N+1)$ quantities.  Let's call the vector ${\bf X}_{N+1} = (X_1, X_2, ..., X_N, X_{N+1})^T$,
 and call its realm matrix then ${\bf R}({\bf X}_{N+1})$.  In general
it will look something like the realm matrix we have just constructed for various aspects of
our gedankenexperiment.  It will have $N+1$ rows, and some number $K$ columns.  Just as an
example, the realm matrix we have already constructed happens to have $(N+1)=16$ rows and $K=16$ columns.
(Mind you, we have not yet specified expectation for the first N components of the quantity
vector to which this realm applies, but let's not let that deter us. I am merely suggesting here an
example of a realm matrix that could be considered to have $(N+1)$ rows.  Let's continue
with the general abstract specification.)\\

Now any such vector of quantities can be expressed as the product of its realm matrix with
a particular vector of events.  The matrix equation, displayed in a form that partitions the
 final row, would look like this:\vspace{.2cm}\\
{\small
\noindent $\left(\begin{array}{c}
X_1 \\
 X_2 \\
  X_3\\
   \hspace*{.2cm}^{\bf .} \\
   \hspace*{.2cm}^{\bf .} \\
   \hspace*{.2cm}^{\bf .} \\
     X_N\\
    *** \\
  X_{N+1}\vspace{.2cm}
       \end{array}\right) \  = \
  \left(\begin{array}{cccccc}
x_{1,1} & x_{1,2} &  \hspace{.2cm}^{\bf .} & \hspace{.2cm}^{\bf .} & x_{1,(K-1)} & x_{1,K} \\
x_{2,1} & x_{2,2} &  \hspace{.2cm}^{\bf .} & \hspace{.2cm}^{\bf .} & x_{2,(K-1)} & x_{2,K} \\
x_{3,1} & x_{3,2} & \hspace{.2cm}^{\bf .} & \hspace{.2cm}^{\bf .} & x_{3,(K-1)} & x_{3,K} \\
\hspace*{.2cm}^{\bf .} & \hspace{.2cm}^{\bf .} & \hspace{.2cm}^{\bf .} & \hspace{.2cm}^{\bf .} & \hspace{.2cm}^{\bf .} & \hspace{.2cm}^{\bf .} \\
\hspace*{.2cm}^{\bf .} & \hspace{.2cm}^{\bf .} & \hspace{.2cm}^{\bf .} & \hspace{.2cm}^{\bf .} & \hspace{.2cm}^{\bf .} & \hspace{.2cm}^{\bf .} \\
\hspace*{.2cm}^{\bf .} & \hspace{.2cm}^{\bf .} & \hspace{.2cm}^{\bf .} & \hspace{.2cm}^{\bf .} & \hspace{.2cm}^{\bf .} & \hspace{.2cm}^{\bf .} \\
x_{N,1} & x_{N,2} & \hspace{.2cm}^{\bf .} & \hspace{.2cm}^{\bf .} & x_{N,(K-1)} & x_{N,K} \\
*** & ** & ** & ** & ** & ** \\
x_{(N+1),1} & x_{(N+1),2} & \hspace{.2cm}^{\bf .} & \hspace{.2cm}^{\bf .} & x_{(N+1),(K-1)} & x_{(N+1),K}
\end{array}\right) \
\left(\begin{array}{c}
({\bf X}_{N+1} = {\bf x}_{^.1}) \\
({\bf X}_{N+1} = {\bf x}_{^.2}) \\
({\bf X}_{N+1} = {\bf x}_{^.3})\\
   \hspace*{.2cm}^{\bf .} \\
   \hspace*{.2cm}^{\bf .} \\
   \hspace*{.2cm}^{\bf .} \\
   \hspace*{.2cm}^{\bf .} \\
({\bf X}_{N+1} = {\bf x}_{^.(K-1)})\\
 ({\bf X}_{N+1} = {\bf x}_{^.K})
       \end{array}\right).$
}\vspace{.2cm}\\
\noindent  On the left of this equation is the column vector of the quantity observations under consideration.
To the right of the equality comes firstly the $(N+1) \times K$ realm matrix whose $K$ columns list all the possible columns of numbers that
could possibly result as the observation vector.  These $K$ columns, each of which has $(N+1)$ components, correspond to vectors denoted as
${\bf x}_{^.1}, {\bf x}_{^.2}, {\bf x}_{^.3}, ..., {\bf x}_{^.(K-1)}$, and ${\bf x}_{^.K}$.
(The initial subscripted dot denotes that this is a whole column of numbers. The number that follows the dot denotes
which of the columns of the matrix it is we are talking about.)  This matrix is multiplied by the
final $K \times 1$ column vector of
events that identify whether the quantity vector ${\bf X}_{N+1}$ turns out upon observation
to be the first, the second, ..., or the $K^{th}$
of these listed columns. We shall denote this vector
by  ${\bf Q}({\bf X}_{N+1})$, and call it
``the partition vector generated
by ${\bf X}_{N+1}$''.   One and only one of its component events will equal $1$ and the rest will equal $0$.  But we do
not know which of them is the $1$, because we do not know which column of possibilities in the realm matrix
will be the one that
represents the observed outcome of the vector of quantities ${\bf X}_{N+1}$.\\

We can represent this matrix equation more concisely and in a useful form
by writing it in an abbreviated partitioned form:\vspace{.2cm}

\hspace*{3.8cm}$\left(\begin{array}{c}
 {\bf X}_N\\
 X_{N+1}
       \end{array}\right) \  \ = \ \ {\bf R}\left(\begin{array}{c}
 {\bf X}_N\\
 X_{N+1}
       \end{array}\right)\vspace{.3cm} \ \  {\bf Q}({\bf X}_{(N+1)})$ \ .\\

 The payoff from constructing this matrix structure is that now every row of this partitioned equation has
 on its left-hand side the unknown value of a quantity, $X_i$.  On the right-hand side in that row appears a
 list of the possible values of that quantity, each multiplied in a linear combination with the  events that
 denote whether each of them
 is indeed the value of this quantity (in the context of the observed values of the other quantities shown in
 that column as well).
 Each row of this equation specifies how a different one of the quantities under consideration equals a linear
 combination of events.  We have heard of that before.  The expectation of a linear combination equals the
 same linear combination of expectations for those events, which would be their probabilities if we could
 specify values for them.  This tells us that
 we can evaluate an expectation operator on this partitioned equation to yield the result that \vspace{.2cm}

 \hspace*{4cm}$E\left(\begin{array}{c}
 {\bf X}_N\\
 X_{N+1}
       \end{array}\right) \  \ = \ \ {\bf R}\left(\begin{array}{c}
 {\bf X}_N\\
 X_{N+1}
       \end{array}\right)\vspace{.3cm} \ \  P[{\bf Q}({\bf X}_{(N+1)})]$ \ .\\

 Well, we have not mentioned anything about probability specifications appearing in the vector
 $P[{\bf Q}({\bf X}_{(N+1)})]$  on the right-hand side of this equality.  The only restrictions of
  probability are that these must be non-negative numbers that sum to $1$, since the vector ${\bf Q}({\bf X}_{(N+1)})$
  constitutes a partition.  We have mentioned only that
 expectations have been identified for the first $N$ components of the vector on the left-hand side, $E({\bf X}_N)$.
 Yet we can compute something important on the basis of this realization.
 The linearity of this equation ensures that the implied value
 for the expectation of the final unspecified component $E(X_{N+1})$ must lie within a specific interval.
 It is computable as the \vspace{.2cm} \\
 \noindent \hspace*{4.3cm}minimum
 and maximum values of ${\bf R}(X_{N+1}) {\bf q}_K$ \vspace{.2cm} \\
 \noindent \hspace*{3.55cm}subject to the linear restrictions  \ \  ${\bf R}({\bf X}_{N}) {\bf q}_K = E({\bf X}_N)$\ \ ,\vspace{.2cm} \\
 \noindent \hspace*{2.5cm}as required of the
 expectations that we have presumed to be specified, \vspace{.2cm} \\
\noindent \hspace*{2cm}and where the components of ${\bf q}_K$ must be non-negative and must sum to $1$.\\

 Such a computation is provided by the procedures of a linear programming problem.  The ``solutions'' to these
 linear programming problems are the vectors ${\bf q}_{min}$ and ${\bf q}_{max}$ that yield these minimum
 and maximum values for $E(X_{N+1})$ subject to these constraints.  The final row vector identifying
 $E(X_{N+1})$ whose extreme values
 we seek is called ``the objective function'' of the problems.  Its coefficients are the partitioned final row of the
 general realm matrix we identify as ${\bf R}(X_{N+1})$.  Notice that that $X$ is not bold.  It represents merely the final
quantity in the column vector ${\bf X}_{N+1}$.  The coefficients vector of the objective function is the final
 row vector of the realm matrix.\\

 Here are the specific details appropriate to our gedankenexperiment.\\

\noindent \hspace*{2.5cm} ${\bf E}\left(\begin{array}{c}
                1 \\
    A({\bf a})B({\bf b}) \\
    A({\bf a})B({\bf b'}) \\
    A({\bf a}')B({\bf b}) \\
    A({\bf a}')B({\bf b}')
       \end{array}\right) \ \ \ \ = \ \ \ \
  \left(\begin{array}{cccccccc}
1 & 1 & 1 & 1 & 1 & 1 & 1 & 1 \\
1 & 1 & 1 & 1 & \llap{$-$}1 & \llap{$-$}1 & \llap{$-$}1 & \llap{$-$}1 \\
1 & \llap{$-$}1 & 1 & \llap{$-$}1 & \llap{$-$}1 & 1 & \llap{$-$}1 & 1 \\
1 & 1 & \llap{$-$}1 & \llap{$-$}1 & 1 & 1 & \llap{$-$}1 & \llap{$-$}1 \\
1 & \llap{$-$}1 & \llap{$-$}1 & 1 & 1 & \llap{$-$}1 & \llap{$-$}1 & 1
\end{array}\right) \ \ \ {\bf q}_{8}$ \vspace{.3cm}\hspace{2.3cm}$(12)$

I have listed the order of the quantities in the vector at left to begin with the sure quantity, $1$, which
equals $1$ no matter what happens in the gedankenexperiment.  There follow the four summands of the CHSH quantity
$s$, of which we have noticed that each one of them is restricted in the gedankenexperiment to equal a function
value of the other three. That is why there are only eight columns in their realm matrix,
 as opposed to sixteen columns in the expansive
realm matrix we have already examined.  As to the components of the vector ${\bf q}_8$ at the right of
the right-hand side,
notice that quantum theory says nothing at all about these, individually.
Each of them would equal the probability that the 4-ply gedankenexperiment
would yield detection products designated by a specific column of the realm matrix.
However, these would involve the joint detection of photon products in four distinct measurements that are known to
be incompatible.  On account of the generalised uncertainty principle, quantum theory eschews specification
of such probabilities.  Nonetheless, for any individual photon detection product in a specific experimental design,
denoted on the left-hand side of the equation,
quantum theory does specifies an expectation value of either $1/\sqrt{2}$ or $-1/\sqrt{2}$, as we have recognized.
Since these four products are not all free to equal $+1$ or $-1$ at the same time, we may assert expectation
values for any three of them, and use linear programming computations to find the cohering bounds on the
expectation of the fourth that would accompany them, yielding bounds on the expectation equations $(11)$.

\subsection{The result:  quantum theory identifies restrictions on the valuation of ${\bf q}_8$}

This is what we find.  The columns of the matrix below display the computed results of the paired
${\bf q}_{min}$ and ${\bf q}_{max}$ vectors corresponding to four linear programming problems.  Each
of them determines a bound on an expected function value that appears in one of the four
forms of the expectation value $E[s(\lambda)]$ which
we displayed in equation $(11)$.  The first pair of columns, for example, identify
the fifth row of the matrix in equation $(12)$ as the objective function, $E[A({\bf a'},{\bf b'})]$,
constrained by QM-specified
values of the expectations of the first four rows.  The second pair of columns identify the fourth
row of $(12)$ as the objective function constrained by QM specifications of expectations for
rows $1,2,3,$ and $5$, and display the appropriate solution vectors; and so on.\\

\noindent {\small $\left(
  \begin{array}{ccccccccc}
 & _{min}({\bf a'},{\bf b'}) & _{max}({\bf a'},{\bf b'}) & _{min}({\bf a'},{\bf b}) & _{max}({\bf a'},{\bf b}) & _{min}({\bf a},{\bf b'}) & _{max}({\bf a},{\bf b'}) & _{min}({\bf a},{\bf b}) & _{max}({\bf a},{\bf b}) \\
q_1 &          0  &  0.1464   & 0 &  0.1464  &  0.5607  &  0.7803   &      0  &  0.1464 \\
q_2 &    0.7803   & 0.5607    &     0  &  0.1464 &   0.1464    &    0    &     0  &  0.1464 \\
q_3 &    0.0732    &     0   & 0.0732    &     0     &    0  &  0.0732     &    0   &      0 \\
q_4 &         0  &  0.1464   & 0.7803  &  0.5607  &  0.1464    &     0  &       0  &  0.1464 \\
q_5 &         0  &  0.1464   &      0  &  0.1464  &  0.1464    &     0  &  0.7803  &  0.5607 \\
q_6 &    0.0732  &       0    &     0  &       0   &      0  &  0.0732  &  0.0732   &      0 \\
q_7 &    0.0732   &      0   &  0.0732 &        0  &       0  &       0  &  0.0732   &      0  \\
q_8 &         0    &     0   & 0.0732   &      0    &     0   &   0.0732  &  0.0732   &      0
  \end{array}
\right)$}\vspace{.3cm}

\noindent Each of these column vectors resides in 8-dimensional space, providing a coherent assessment
of probabilities for the constituent event vector ${\bf Q}({\bf X}_8)$, without specifying precise
probabilities for any of them.  In fact, quantum theory denies itself the capability of identifying
such probabilities precisely.  We will discuss this feature further, below.  However the results of
 the linear programming computations can and do specify
possibilities for what might be specified in a way that would cohere with what quantum theory can and
does tell us.  The columns of this matrix identify some of them.  In fact, these
columns display extreme values of
what are possible.  Any convex (linear) combination of them would cohere with quantum theory as well.
Thus, geometrically the columns constitute vertices of a polytope of quantum-theory-supported possibilities
for $P[{\bf Q}({\bf X}_8)]$.  This polytope is called ``the convex hull'' of these vectors.
However, although we have found eight of them, the rank of the
matrix of all of them is only four!  That is, these eight-dimensional vectors all reside within a
four-dimensional subspace of a unit-simplex.  Why is quantum theory not more specific in specifying the expectation of Bell's
quantity $E(s)$?  We shall delay this discussion until
we have clarified what we have learned from these results of ${\bf q}_{min}({\bf a}^*,{\bf b}^*)$ and
${\bf q}_{max}({\bf a}^*,{\bf b}^*)$.

\subsection{Implied bounds on expected detection products and on $E(s) \in (1.1213, 2]$}

\noindent   According to the prescription of equation $(12)$, each of these
${\bf q}_8$ vectors appearing in Section $7.1$ would identify a vertex of another polytope of cohering
expectation vectors for the components of the CHSH quantity $s$.  Followed at bottom by the implied
expectation $E(s)$, these are\\

{\small
\indent $\left(
  \begin{array}{ccccccccc}
E[A({\bf a})B({\bf b})] &    0.7071  &  0.7071  &  0.7071  &  0.7071  &  0.7071   & 0.7071  & \llap{$-$}1.0000  & \llap{$-$}0.1213\\
E[A({\bf a})B({\bf b'})] &   \llap{$-$}0.7071  & \llap{$-$}0.7071  & \llap{$-$}0.7071  & \llap{$-$}0.7071  &  0.1213   & 1.0000  & \llap{$-$}0.7071  & \llap{$-$}0.7071\\
E[A({\bf a'})B({\bf b})] &    0.7071  &  0.7071  &\llap{$-$}1.0000  & \llap{$-$}0.1213  &  0.7071   & 0.7071  &  0.7071  &  0.7071\\
E[A({\bf a'})B( {\bf b}')] &   \llap{$-$}1.0000  & \llap{$-$}0.1213  &  0.7071  &  0.7071  &  0.7071   & 0.7071  &  0.7071  &  0.7071\\
E[s] &   1.1213  &  2.0000  &  1.1213  &  2.0000  &  2.0000  &  1.1213  &  1.1213  &  2.0000$
    $  \end{array}
\right)$\\
}

\noindent In any of these columns appear three values of $E[A({\bf a^*})B({\bf b^*})]$ specifications supported by
quantum theory, and a fourth value which is either a lower bound or upper bound on any cohering expectation for the
fourth. (By the way, $0.7071$ is the value of $1/\sqrt{2}$ to four decimal places.)
At the bottom of the column is the value of $E(s)$ that would correspond to these four.  The vectors of
the four $E[A({\bf a^*})B({\bf b^*})]$ values are the vertices of the four-dimensional space of QM-supported
expectation values of the gedankenexperiment, and the value of $E(s)$ listed at bottom would be a
quantum-theory-permitting assessment of $E(s)$, Bell's quantity.  All of their convex combinations lie within
Bell's reputed bounds of $[-2,+2]$.  There is more to be said about this,
but let us first address the question of why
quantum theory leaves four dimensions of freedom unaccounted for in its prescriptions.

\subsection{Why are there four free dimensions to the QM specification of $E(s)$\ ?}

Let's just get down to it, without any prelude.  Quantum theory specifies precise values
for outcome probabilities of the photon pair detections at any choice of three angle settings of the gedankenexperiment.
Consider the QM identifications of detection probabilities at
the angles $({\bf a},{\bf b}), ({\bf a},{\bf b'})$, and $({\bf a'},{\bf b})$.
These have been identified in our equations $(1)$, and the corresponding expectations of the detection products
have appeared in equation $(2)$.  If quantum theory were to specify a complete distribution for the outcome of
this gedankenexperiment, it would have to specify eight probabilities.  These would involve three corresponding
to  detection events at any one of the polarizations angles, also jointly at any two of the three detection angles,
 and also at all three of the detection angles.  But according to the uncertainty principle discussed in Section 3,
 the theory eschews commitments regarding the latter four of these probabilities: neither \vspace{.2cm} \\
\indent $P\{[(A({\bf a})=+1)(B({\bf b})=+1)][(A({\bf a})=+1)(B({\bf b'})=+1)]\}$\vspace{.2cm}\ , nor\\
\hspace*{2cm}$P\{[(A({\bf a})=+1)(B({\bf b})=+1)][(A({\bf a'})=+1)(B({\bf b'})=+1)]\}\vspace{.2cm}$\ , nor\\
\hspace*{4cm}$P\{[(A({\bf a'})=+1)(B({\bf b})=+1)][(A({\bf a'})=+1)(B({\bf b'})=+1)]\}\vspace{.2cm}$\ , nor \\
\indent $P\{[(A({\bf a})=+1)(B({\bf b})=+1)][(A({\bf a})=+1)(B({\bf b'})=+1)][(A({\bf a'})=+1)(B({\bf b'})=+1)]\}$\vspace{.23cm}.\\
\noindent For each of these would amount to claims regarding the joint outcomes of incompatible measurements,
characterised by Hadamard matrix operators that do not commute.  Quantum theory
explicitly avoids such claims.  That leaves four dimensions of the eight-dimensional pmf over the four
detection products unspecified ... explicitly!  That is why quantum theory allows four unspecified dimensions
to the expectations it provides regarding the four polarization products on the same pair of photons.\\

Perhaps this comment does need a little bit more explication.  You will need to view equation $(12)$
while reading the following remarks.  They concern assertions that quantum theory does allow us to make,
and those that it doesn't.
Recall that we are considering a linear programming problem in which quantum expectations are asserted for the polarization
products at the angle settings $({\bf a},{\bf b}), ({\bf a},{\bf b'})$, and $({\bf a'},{\bf b})$, and investigating
coherent bounds for expectation of the product at the setting $({\bf a'},{\bf b'})$.
Notice firstly that quantum theory does allow us to, and indeed insists that we\vspace{.2cm}
assert\\
\hspace*{3cm} $E[A({\bf a})B({\bf b})] \ \ = \ \ q_1 + q_2 + q_3 +q_4 - q_5 - q_6 - q_7 - q_8 \ \ = \ \ 1/\sqrt{2}$\vspace{.2cm}\\
 Examining the corresponding columns of the realm matrix seen in $(12)$, it is evident that these
 involve assertions regarding the outcomes of $(A({\bf a})B({\bf b}) = +1)$ and $(A({\bf a})B({\bf b}) = -1)$
 irrespective of the values of $A({\bf a})B({\bf b'})$ and $A({\bf a'})B({\bf b})$. For each of these events
 involve an outcome of the product $A({\bf a})B({\bf b})$ summed over all four possible joint
 outcomes of the products
  $A({\bf a})B({\bf b'})$ and $A({\bf a'})B({\bf b})$.  So these latter two
  incompatible observations would be irrelevant
    to the assertion of this expectation.  The same feature
 would pertain to the required assertions of $E[A({\bf a})B({\bf b'})]$ and $E[A({\bf a'})B({\bf b})]$
 which are involved in the first LP problem.
 Neither of these involves any concomitant assertions regarding observations incompatible with them.
On the other hand, an assertion of a probability for the joint occurrence of two pairs of polarization observations, such as
$P\{[(A({\bf a})=+1)(B({\bf b})=+1)][(A({\bf a})=+1)(B({\bf b'})=+1)]\}$ for example, would require
specifications of the sum $q_1 + q_3$. Examining equation $(12)$ makes clear that it is only columns $1$ and $3$
of the matrix in which this joint event is instantiated.  Asserting a specific value for the sum $q_1+q_3$ would necessarily
entail assessments of
joint probabilities for incompatible events.  The same would be true of any of the other three
probabilities regarding joint events for which quantum theory eschews assessment.\\

If one were to claim, as do the reigning proponents of Bell violations, that the probabilities of
quantum theory support the valuation of $E[s(\lambda)] = 2\sqrt{2}$ according to the derivation that
concluded our Section 5, that would be just plain wrong.  Full stop.\\

Our next project is an amusing one, of actually envisaging
the 4-dimensional polytope of quantum probabilities relevant to the
gedankenexperiment.  This will be achieved by passing this 4-dimensional quantum polytope through our
3-dimensional space, so to view it, just as the inhabitants of 2-dimensional space in {\it Flatland} (Abbott,
1884) viewed the
sphere passing through their lower dimensional world.  It suddenly appeared
 as a point, which gradually expanded to  circles of increasing diameter, and then diminished until they
 suddenly disappeared again.  Let's view what we can
of our 4-dimensional quantum polytope in this way.

\subsection{Transforming the expectation polytope into quantum probabilities}

\noindent  The expected photon detection products displayed in Section $7.2$ can be transformed into $P_{++}$ probabilities
by applying the transformation  \  $P_{++}({\bf a}^*,{\bf b}^*) =  [\,E({\bf a}^*,{\bf b}^*) + 1]/4$ of equation $(3)$
to the eight vertices.  This yields the vertices of another polytope in the space of the probability vector
$[P_{++}({\bf a},{\bf b}), P_{++}({\bf a},{\bf b'}), P_{++}({\bf a'},{\bf b}), P_{++}({\bf a'},{\bf b'})]$ displayed below: \\

$\left(
  \begin{array}{ccccccccc}
P_{++}({\bf a},{\bf b}) &   0.4268  &  0.4268   & 0.4268   & 0.4268  &  0.4268  &  0.4268  &       0  &  0.2197\\
P_{++}({\bf a},{\bf b'}) &     0.0732  &  0.0732   & 0.0732   & 0.0732  &  0.2803  &  0.5000  &  0.0732  &  0.0732\\
P_{++}({\bf a'},{\bf b}) &     0.4268  &  0.4268   &      0   & 0.2197  &  0.4268  &  0.4268  &  0.4268  &  0.4268\\
P_{++}({\bf a'},{\bf b'}) &          0  &  0.2197   & 0.4268   & 0.4268  &  0.4268  &  0.4268  &  0.4268  &  0.4268$
    $  \end{array}
\right)\; .$

\subsection{And now viewing it ! ...  as it passes through our space}


The convex hull of the 4-D column vectors shown in Section $7.4$ can be visualized through a
sequence of 3-D intersections it affords with slices perpendicular to any one of its axes.  Figure $3$ 
on the next page 
displays such a sequence, by slices perpendicular to the $P_{++}({\bf a}',{\bf b}')$
axis at values increasing from $0$ to $.4268$.  When $P_{++}({\bf a}',{\bf b}') = 0$, the
intersection of the slice identifies only a single vertex point $(.0732,.4268,.4268)$
which appears in the subplot $(1,1)$.  See also column one of the matrix in Section $7.4$.
As the value of $P_{++}({\bf a}',{\bf b}')$ for the
slice level increases to a .1098 in subplot $(2,1)$, the
intersection appears as a tetrahedron.  The size of the intersecting tetrahedron
increases further at the probability level $.2197$ in subplot $(3,1)$.
 The tetrahedrons continue to increase in size as the level of the $P_{++}({\bf a}',{\bf
b}')$ increases still further to $.2561$ in subplot (1,2),
but a corner of their intersections begins to be cut off there.  This clipped portion is cut
more severely from the enlarging polytope as $P_{++}({\bf a}',{\bf b}')$ increases
further, displayed in subplots $(2,2)$ through $(3,2)$ which is our view of the polytope
when it suddenly disappears.\\

\begin{figure}[!h]
\center
\includegraphics[scale=0.75]{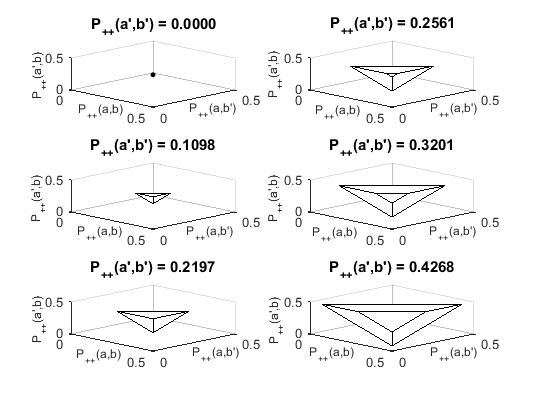}
\caption{Sequential intersections of the 4-D convex hull of vectors $[P_{++}({\bf a},{\bf
b}), P_{++}({\bf a},{\bf b}'),
P_{++}({\bf a}',{\bf b}), P_{++}({\bf a}',{\bf b}')]$ with slices perpendicular to the
$P_{++}({\bf a}',{\bf b}')$ axis, at levels increasing from $0$ to $.4268$.  }
\label{fig:Working4dP++pdt5587}
\end{figure}

The symmetry of the configuration implies that slices along the other axes would create
identical intersection sequences.  This Figure was produced by my colleague
Rachael Tappenden who has also produced a moving sequence of this progression as the
intersections proceed along the  $P_{++}({\bf a}',{\bf b}')$ axis in more refined stages.   I will make this
available online when I get to it.

\section{What to make of Aspect's and subsequent empiricism}

Taken in by the alluring derivation of Section $5.3$ which ignores the symmetric functional relations among the
polarization products of the gedankenexperiment, Aspect and followers were convinced that Bell's inequality has
been defied, and that the theory of hidden
variables must be rejected. This conclusion would support the assertion that quantum theory has identified the
structure of randomness which supposedly inheres in Nature at its finest resolution.  The behaviour of the photons
is considered to be governed purely by a probability distribution.  It remained only to devise some physical
experiments that could verify the defiance of the inequality.  \\

According to the tenets of objective probability theory and its statistical programme, probabilities are not observable
quantities.  What are observable are outcomes of random variables which are generated by them. It is a matter of
statistical theory to devise
methods for estimating the unobservable probabilities and their implied expectations from carefully observed outcomes
of the random variables they generate.  Understood in this way, equation $(8)$ which I repeat here constitutes a structure requiring
estimation if the violation of Bell's inequality is to be verified:\vspace{.2cm}\\
\indent \hspace*{.6cm}$E[s(\lambda)] \, \vspace{.15cm} \ \ = \ \ \,E[A({\bf a})B({\bf b})] \ - \
E[A({\bf a})B({\bf b}')]  \ + \ E[A({\bf a}')B({\bf b})] \ + \
E[A({\bf a}')B({\bf b}')]. \ \ \ \ \ \ \ \ \ $\vspace{.2cm}\\
\noindent According to long respected statistical procedures, the unobservable expectations of
detection products on the right-hand-side of this equation can be estimated by the generally
applicable non-parametric method of moments.  Supported by the probabilistic law of large numbers,
its validity as an estimating procedure stems from the 1930's.\\

    The programme for estimating equation $(8)$ would proceed as follows.  To estimate the first component of $E[s(\lambda)]$, which is $E[A({\bf a})B({\bf b})]$, one would conduct $N$ independent polarization experiments at the angle setting $({\bf a},{\bf b})$, and record the value of the polarization products $A({\bf a})B({\bf b})$ observed in each case, these being either $-1$ or $+1$.  The average of these values would provide a method of moments estimate of the expectation  $E[A({\bf a})B({\bf b})]$ which is common to all of these random experiments.  A similar programme would be followed in estimating the other three components of $E[s(\lambda)]$.\\

Using the notation of Aspect (2002) we would conduct $N$ repetitions of the CHSH/Bell experiment with the relative polarizing angles set at $({\bf a},{\bf b})$, resulting in $N_{++}({\bf a},{\bf
b})$ observations of $(A({\bf a}),B({\bf b})) = (+,+)$, $N_{+-}({\bf a},{\bf
b})$ observations of $(+,-)$, $N_{-+}({\bf a},{\bf
b})$ observations of $(-,+)$, and $N_{--}({\bf a},{\bf
b})$ observations of $(-,-)$.
An estimated version of equation $(8)$ would then be expressed  as \vspace{.2cm}\\
\indent $\displaystyle \hat{E}[s(\lambda)] \, \vspace{.2cm} \ \ \equiv \ \ \,\displaystyle \hat{E}[A({\bf a})B({\bf b})] \ - \
\displaystyle \hat{E}[A({\bf a})B({\bf b}')]  \ + \ \displaystyle \hat{E}[A({\bf a}')B({\bf b})] \ + \
\displaystyle \hat{E}[A({\bf a}')B({\bf b}')]\, , \ \ \ \ \ \ (13)$\vspace{.2cm}\\
where the component estimator $\displaystyle \hat{E}[A({\bf
a})B({\bf b})]$ is defined by \vspace{.3cm}

\hspace*{1.1 cm} $\displaystyle \hat{E}[A({\bf
a})B({\bf b})] \ \equiv \ \frac{[N_{++}({\bf a},{\bf
b})-N_{+-}({\bf a},{\bf b})-N_{-+}({\bf a},{\bf b})+N_{--}({\bf a},{\bf
b})]}{[N_{++}({\bf a},{\bf b})+N_{+-}({\bf a},{\bf b})+N_{-+}({\bf a},{\bf
b})+N_{--}({\bf a},{\bf b})]}\ \ $, \vspace{.2 cm} \hspace{1.7cm}$(14)$ \\
with a similar specification for the components of $\hat{E}[s(\lambda)]$ pertaining to the
relative angles $({\bf a}',{\bf b})$, $({\bf a},{\bf b}')$, and $({\bf a}',{\bf b}')$.
The denominator of $(14)$ is equal to $N$, the number of experiments run at this angle,
merely displayed as the sum of its four component counts of $N_{\pm\pm}({\bf a},{\bf b})$. \\

The momentous results were published in Aspect et al (1982), confirming the apparent defiance of
Bell's inequality to several decimal places.

\subsection{Examining and reassessing Aspect's empirical results}

What are we to make of Aspect's and subsequent empirical results$\,$? \\

Aspect (2002, page 15, and 1982) reports the estimation $\hat{E}[s(\lambda)]$ from experimental data,
using the method of
moments as  defined in equations
$(13)$ and $(14)$. Of course actually, it is impossible to conduct an experiment on a single pair of
photons at all four angle settings, much less conduct a sequence of such experiments.
Instead,  experimental sequences of observations using
different photon pairs were generated at each of four angle settings.  These were presumed
to provide independent estimates of the four expectations as they appear in equation $(14)$.
These independent estimates were then inserted into equation $(13)$, yielding Aspect's touted
estimate  $\hat{E}[s(\lambda)]$ near to $2\sqrt{2}$. \\

Although experimentation protocols have subsequently been improved to account for
the challenges of possible loopholes during the following thirty years, the estimation
procedures using the improved data have been the same.  Results from several of the
improved protocols have been reported only in the form of so-called p-values of significance
for hypothesis tests posed as to whether $E[s(\lambda)]$ exceeds $2$ or not.  The results
have been lionized, apparently quite impressive, and deemed to be decisive. \\

We can now recognize the fault in Aspect's estimation procedure
which allows complete liberty in all
four polarization product estimations $\hat{E}[A({\bf a}^*)B({\bf b}^*)]$, using experimental
incidence values of $N_{\pm\pm}({\bf a}^*,{\bf b}^*)$ from many experimental runs {\it
with different photon pairs}.
Each of his experimental observations may be whatever value it happens to be at its
experimental angle setting, identifying whatever value of polarization product that it
does.
However, if the estimation were meant to apply to the ontological understanding of
$s(\lambda)$ in the gedankenexperiment within which he and Bell couch their theoretical
claims, he would have to adjust this methodology.  One might well pick experimental runs using
three different photon pairs at any three angles one
wishes, to simulate the behaviours $A({\bf a}^*)B({\bf b}^*)$ for any three
polarization products of a single pair of photons.  However, to be consistent with the
Aspect/Bell problem as posed  for this single pair of photons at all four relative
angle settings, one then would need to compute the implied value of the polarization
product observation for the fourth angle according to the functional form that we have
identified in equation $(10)$.  The same functional form connects the detection product at
any one of the four angle settings to the other three.\\

Statistical estimation values reported by Aspect as well as those by
subsequent research groups over
 the past thirty years have no relevance to the estimation of
$E[s(\lambda)]$ as it is understood to pertain to four spin products on a single pair of
photons.  It is perfectly reasonable to find estimation values exceeding the bounds of $[-2,+2]$ as
they have.  For although these results {\it could reasonably pertain} to an estimate of
$E[s(\lambda)]$ with
$s(\lambda)$ defined as a combination of polarization products on {\it four different
pairs of photons}, they {\it do not pertain} to Bell's inequality which is relevant to
a 4-ply gedankenexperiment on the same pair of photons at all four angle settings.
In the context to which their experimental results are appropriate,
$E[s(\lambda)]$ is not bound by the Bell bounds of
$[-2,\,+2]$, but rather by the interval $[-4,\,+4]$ which is unchallenged in this context.\\

Nonetheless, Aspect's empirical estimation programme might be adjusted to account for the
symmetric functional relations that would necessarily characterise the imagined results of
the gedankenexperiment.  In the next subsection I shall display the unsurprising results
of such an adjusted methodology.  They do not suggest any defiance of Bell's inequality at all.
The simulation I construct will mimic the way Aspect's data needs to
be treated, recognizing his data as the result of conditionally independent experiments on distinct pairs
of photons at each of the four relative angle settings of the polarizers.

\subsection{Exposition by simulation}

Because Aspect's experimental observation data is not available in full, a method for correcting
his estimation procedure shall now be displayed, along with its numerical implications.  To
begin, four columns of one million $(10^6)$ pseudo random numbers, uniform on $[0,1]$,
were generated with a MATLAB routine.  These were then transformed into simulated
observations of paired photon polarization experiments at the four
relative angles we have been
studying.  These transformations were performed using the QM probabilities based on
calculations of $\frac{1}{2}\;cos^2({\bf a}^*,{\bf b}^*)$ and $\frac{1}{2}\;sin^2({\bf
a}^*,{\bf b}^*)$ as described in our equations $(1)$.   Each resulting
simulated polarization pair was then multiplied together to yield a polarization product.
 In this way were created four columns of simulated observations corresponding to polarization
products from one million experiments at each of the four angles: \
 $({\bf a}',{\bf b}'),\;({\bf
a},{\bf b}'),\;({\bf a},{\bf b}),\;({\bf a}',{\bf b})$.  We shall refer to this matrix
of simulated polarization products below as the SIMPROD matrix.\\

Aspect's estimation equation $(14)$ was applied to each of these columns, yielding
estimates of the expected polarization product pertinent to that column, $\hat{E}[A({\bf a}^*)B({\bf b}^*)]$.
These appear in the first row of Table $1$.
These four estimates were then inserted into equation $(13)$ appropriately to yield an
Aspect estimate $\hat{E}[s(\lambda)] = 2.827738$, appearing in the second row of the
Table under {\it each} of these columns. This number is quite near to $2\sqrt{2} \approx
2.828427$, as was Aspect's reported empirical estimate, proposed as an evidential  violation of
Bell's inequality.  As we now know, the problem is that when the product observations are
supposed to apply to {\it the same} photon pair, the observed value of the polarization product at
any angle is required to be related to the product at the other three angles via the
functional equation we exemplified our equation $(10)$. The four of them may not all
range freely, as they may in
real experiments on {\it different} pairs of photons.  Rather, they are required to be bound
by the symmetric functional relation $G(^.,^.,^.)$ that we have identified.  The rows of the matrix
SIMPROD do {\it not} respect this requirement, so the Aspect estimate $\hat{E}[s(\lambda)]$ which
they produce cannot be used to estimate the expected value of $s(\lambda)$ for the gedankenexperiment.
We shall now endeavor to correct this error.
\begin{center}
\noindent {\bf Table 1:  Corrections to Aspect's estimate of
$E[s(\lambda)]$}\vspace{.2cm}\\
\noindent  \begin{tabular}{c c c c c}
$ ({\bf a}^*,{\bf b}^*)$ & $ ({\bf a},{\bf b})$ & $({\bf a},{\bf b}')$ & $ ({\bf a}',{\bf b})$ & $({\bf
a}',{\bf b}')$    \\ 
$\hat{E}[A({\bf a}^*)B({\bf b}^*)]$  &  $0.707232 $ & \llap{$-$}$0.706186 $ & $  0.706840
 $ & $ 0.707480$\\
Aspect\ $\hat{E}[s]$ &   $ 2.827738 $ & $  2.827738 $ & $  2.827738  $ & $  2.827738$ \\
Functional\ $\hat{E}[A({\bf a}^*)B({\bf b}^*)]$\ \ \   &  \llap{$-$}$0.353078 $ & $   0.354348
$ & \llap{$-$}$  0.354766 $ & \llap{$-$}$0.353934  $\\
Corrected\ $\hat{E}[s]$ &   $    1.767180 $ & $   1.767204  $ & $  1.765740 $ & $
1.766964 $ \\
   \end{tabular}\\
 \end{center}

\indent The third row of Table $1$ has been generated then by first applying the function $G(^.,^.,^.)$
to each choice of three components of the rows of the SIMPROD matrix.  Each result was entered into
the same row of a companion matrix of the same size, but placed into the column corresponding to the
column entry that was {\it not} used in the evaluation of the $G$ function.  Let's call this matrix
by the name SIMGEN.  Next, Aspect's estimation equation $(14)$
was applied to each of the four columns of SIMGEN,
 and the result is printed in the third row of Table $1$, labeled
 ``Functional\ $\hat{E}[A({\bf a}^*)B({\bf b}^*)]$''.  These display estimates of $E[G(^.,^.,^.)]$
 required for estimation of the four alternative expectation equations $(11)$.
 In this way we can be considered to have
generated $4$ times $10^6$ simulated versions of the Aspect/Bell
gedankenexperiment.  Their component results can be taken to be any
three simulation results from a row of SIMPROD along with the fourth result being the functionally
generated result found in the same row and the appropriate fourth column of SIMGEN.
Finally, the last row of Table $1$ presents the estimated values of $E[s(\lambda)]$ deriving from
these simulated experiments.  They appear as ``corrected estimates'', column by column, for each of
which the $\hat{E}[G(^.,^.,^.)]$ is the one appropriate
to that column while the other three expected polarization products are those appropriate to the
other three columns of row $1$ of the Table.  The elements of this row display corrected estimates of $E[s(\lambda)]$
as they should be calculated with the simulated Aspect data.   Each of these four estimates is slightly
different from the others.  Averaging them
over the four ways of
generating a column of polarization products from the other three columns of simulated
products would yield a ``Corrected estimate'' of $E[s(\lambda)]$ as $1.766772\,$, well
within the Bell bounds of $[-2,\,+2]$.\\

Based on Aspect's report of his experimental data, I feel quite sure that applying this same
estimation procedure to his experimental data, considered as a simulation of the impossible
gedankenexperiment, would yield a similar result.\\

Results on the order of this peculiar number are quite stable over repeated runs of this
simulation as described.  Since the theoretical analysis reported in this article yields
only an interval of cohering possibilities for $E[s(\lambda)]$, this simulation leaves us
with a tantalizing problem of how to account for this stable result, which is quite near
to $[3/\sqrt{2} - 1/(2\sqrt{2})] \approx 1.767766952966369$.  Most likely, this specific result is  a
construct of the independence feature embedded in the simulation results across angle pairings.  Such
a feature  would be highly suspect in Nature, given what we know now about quantum
entanglement itself in a single experiment.
I should mention that among all distributions in the polytope cohering with the prescriptions of
quantum theory, the maximum entropy distribution inheres an expectation value $E(s) = 1.1522$.  Discussion
of its assessment and related issues must await another forum.
However, there can be no real empirical
evidence on the issue, since it is impossible in principle to activate the setup of the four imagined
simultaneous  experiments on a single pair of photons.  Thus, the physicists' long interest in
the fabled gedankenexperiment.

\subsection{A comment on empirical work and statistical estimation}

While Aspect's conception of statistical estimates appropriate to the photon detection
problem is understandable, and corrections can be made to improve its relevance to the
Aspect/Bell problem, developments of statistical theory and practice during the past
fifty years have surely generated superior methods for evaluating  the physical theory of
quantum behavior. These rely on the subjective theory of probability which, under the
  leadership of Bruno de Finetti and researchers adhering to his viewpoint, has gained
substantial credibility from the past half-century of research in the foundations of
probability and statistics.  There are even some prominent physicists among its proponents,
though not many.  Proclaimers of inherent randomness in the physics of quantum behaviour
have won the day for now, largely on the basis of the mistaken violation of Bell's inequality
that we have debunked in this essay.  In the very least, it is apparent that calls
  for open access
to raw data (Khrennikov, 2015) from several well-known research programs that publish
summary results, usually in the form of p-values, need to be heeded.

\section{Concluding comments }

The mathematical structure of the Aspect/Bell problem and its resolution align well
 with the theory of subjective probability.  This viewpoint is in keeping with Einstein's
interpretation of quantum mechanics, known by his famous adage that the old one does not
roll dice. However, readers more comfortable with the standard realist interpretation of quantum
mechanics may also consider the probabilities as ontic properties of the photons themselves
without disturbing the mathematical issues we have engaged.
Anyone who professes uncertain knowledge about the possible values of a quantum optical
gedankenexperiment may assert whatever probabilities are deemed appropriate for the sixteen
possible observation vectors displayed in block one of our experimental realm matrix.
This may involve as many or as few expectations as one
wishes, and whether based on the theory of
quantum mechanics or not.  These of course need to be assessed
scientifically in the light of what evidence can be brought to bear.
Similarly, realist proponents of quantum theory may hypothesize whatever
probability values they think it prescribes.  However, since the sixteen vectors of
possible polarization observations listed in the realm matrix provide an exclusive
and exhaustive list of possible gedankenexperiment results, the sum of these probabilities must equal $1$ for
anyone who makes
coherent assertions.  This understanding is what resolves the conundrum posed by apparent
violations of Bell's inequality.\\

As to the characterization of the theory of hidden variables, this is another endeavour
that has been misconstrued in accepted literature, largely on the basis of the mistaken understanding
of the defiance of Bell's inequality which we have corrected here.  I have examined this matter in a
separate article,
entitled ``Resurrection of the principle of local realism and the prospects for supplementary variables.''
For now I shall merely state that mathematically, the theory of supplementary variables specifies the form of a mixing
density $\rho(\lambda)$ that can be made isomorphic to {\it any} coherent distribution over the empirical
observations of polarization experiments whatsoever.  It matters not whether they are the prescriptions
of quantum theory or not.
At any rate, no coherent distribution over observable quantities, whether considered to be a
formalization of hidden variables theory or not,
supports the defiance of Bell's inequality.\\

Virtually all discussion of quantum probabilities since the original work of Bell
has supported the
conclusion  that probabilities pertinent to quantum behaviour can violate the
seemingly innocuous inequality that he identified.  The mathematical error that has been
discovered and reported here substantiates the end of an era of accepting this
conclusion.  The results we have aired will have ramifications for many published estimations
based on more sophisticated experimentation as well.  There are further consequences for a
host of theoretical issues
that have been studied and discussed in the context of a mistaken understanding.  These
include related notions of hidden variables, entangled particles, and information
transfer.  Discussion of these topics do require philosophical attention to a variety of
conceptual constructs in which they are imbedded. However, the analysis of Aspect/Bell
presented here has nothing to do with philosophical distinctions.  It has identified a
mathematical error in accepted work that must be recognized no matter what might be the
philosophical positions of interested parties.  Probabilistic forecasts motivated by
quantum theory do not violate any laws of probability theory.  Full stop. \\

Discussions of related issues proceeding henceforth will need to begin with this new
recognition.  Interestingly, this resolution was suspected in some way by
Bell himself, though not the analytical detail.  This was clearly evident in his musings
on the hidden variables question in Bell (1971) which he himself had reprinted in a
collection of his publications, Bell (1987).  I have recently learned of an article of 
Adenier (2001) which seems to have been printed only by arXiv.  It recognises quite clearly 
the relevance of Bell's inequality exclusively to a gedankenexperiment on the same pair of 
photons at all four relative polarization angles.  My discovery of the functional relations 
involved and the 4-D polytope of cohering quantum theoretic distributions for the 
components of the 4-ply quantity $S$ are truly novel.\\

A final reference relevant to this analysis  is the article of Romano
Scozzafava (2000) on the role of probability in statistical physics. He discusses several issues
that clarify fundamental matters in the context of the constructive mathematics of Bruno de
Finetti's operational subjective statistical method.

\section*{Acknowledgments}
Thanks especially to Mike Ulrey for stimulating me to work on this problem, and for very
helpful discussions;  also to Duncan Foley who has been a keen supporter of my investigations over the
 years, with many helpful comments.  Neither of them should be presumed to concur with every aspect
 of my analysis.   Thanks also to my colleague Rachael
Tappenden for programming the 3-D slices of the QM-motivated coherent prevision polytope using MATLAB,
and for discussions concerning the linear algebraic structure of the problem.  The University of
Canterbury provided computing and research facilities.  Thanks to Paul Brouwers, Steve
Gourdie, and Allen Witt for IT service and consultation.

\section*{References}

 \noindent {\bf Adeneir, G.} (2001) A refutation of Bell's theorem, arXiv:quant-ph/0006014v3.\vspace{2mm}\\
\noindent {\bf Aspect, A.} (2002) Bell's Theorem: the naive view of an experimentalist,
{\it Quantum [Un]speakables \\
\indent -- from Bell to Quantum Information}, R.A. Bertlmann and A. Zeilinger (eds.),
Springer.\vspace{2mm} \\
\noindent {\bf Aspect, A., Grangier, P., and G\'{e}rard R} (1981)  Experimental tests of realistic local \\
\indent theories via Bell's theorem, {\it Physical Revue Letters} {\bf 47} (8): 460–3.\vspace{2mm} \\
\noindent {\bf Aspect, A., Dalibard, J., and G\'{e}rard R} (1982)  Experimental tests of Bell's inequalities \\
\indent using time-Varying analyzers, {\it Physical Revue Letters} {\bf 49} (25): 1804-7.\vspace{2mm} \\
\noindent {\bf Bell, J.S.} (1964) On the Einstein Podolsky Rosen paradox, {\it
Physics},{\bf 1}(3), 195-200.\vspace{2mm} \\
\noindent {\bf Bell, J.S.} (1966) On the problem of hidden variables in quantum
mechanics, {\it Reviews of Modern} \\
\indent {\it Physics}, {\bf 38}(3), 447-452.\vspace{2mm} \\
\noindent {\bf Bell, J.S.} (1971) Introduction to the hidden variables question,
reprinted in  Bell, J.S. (1987),\\
\indent 29-39.\vspace{2mm} \\
\noindent {\bf Bell, J.S.} (1987) {\it Speakable and Unspeakable in Quantum Mechanics},
Cambridge Univ Press.\vspace{2mm} \\
\noindent {\bf Clauser, J.F., Horne, M.A., Shimony, A., and Holt, R.A.} (1969) Proposed
experiment \\
\indent to test local hidden-variable theories, {\it Physical Review Letters},  {\bf 23},
 880.\vspace{2mm} \\
\noindent {\bf Einstein, A., Podolsky, B, and Rosen, N.} (1935) Can quantum mechanical
description of
\indent physical reality
be considered complete? {\it The Physical Review}, {\bf 47}, 777-780.  \vspace{2mm} \\
\noindent {\bf de Finetti, B.} (1974, 1975) {\it Theory of Probability}, 2 volumes,
A.M.F. Smith and A. Machi  \\
\indent (trs.),  New York: John Wiley.\vspace{2mm} \\
\noindent {\bf Khrennikov, A.} (2015) Unuploaded experiments have no result, arXiv
1505.04293v2. \vspace{2mm} \\
\noindent {\bf Lad, F.} (1996) {\it Operational Subjective Statistical Methods: a
mathematical, philosophical and \\
\indent historical introduction},  New York: John Wiley.\vspace{2mm} \\
\noindent {\bf Lad, F., Dickey, J.M., and Rahman, M.} (1990) The fundamental theorem of
prevision,
\indent {\it Statistica},
{\bf 50}, 19-38. \vspace{2mm} \\
\noindent {\bf Scozzafava, R.} (2000) The role of probability in statistical physics,
{\it Transport Theory and Stat-} \\
\indent {\it istical Physics}, {\bf 29}, 107-123.\vspace{2mm} \\
\noindent {\bf Whittle, P.} (1970) {\it Probability},  Harmondsworth, Middlesex:  Penguin.\vspace{2mm} \\
\noindent {\bf Whittle, P.} (1971) {\it Optimization under Constraints},  London: Wiley.

\end{document}